\documentclass[12pt]{aastex62}

\begin{document}

\parindent=1.0cm

\title{THE ENVIRONMENTS AROUND W SERPENTIS SYSTEMS: INDEPENDENT LIMITS 
ON SYSTEM MASSES AND EXTENDED ENVELOPES}

\author{T.J. Davidge}

\affiliation{Dominion Astrophysical Observatory,
\\Herzberg Astronomy \& Astrophysics Research Center,
\\National Research Council of Canada, 5071 West Saanich Road,
\\Victoria, BC Canada V9E 2E7\\tim.davidge@nrc.ca; tdavidge1450@gmail.com}

\begin{abstract}

	Information extracted from the 
GAIA Data Release 3 is used to examine the stellar contents 
within projected separations of 10 parsecs 
from eight close binary systems that are either classical W Serpentis 
systems or related objects. The goal is to search for remnant star 
clusters or moving groups with proper motions that are similar to those of 
the binaries. While some of the binary systems have proper motions that 
are distinct from those of the majority of stars within the search area, 
there is still a tendency for W Ser stars to be accompanied by companions 
with separations on parsec or larger scales. At least three candidate 
companions are identified within the search area for 
each system, although in the majority of cases the numbers are much higher. 
Evidence is presented that SX Cas is near the center of a diffuse cluster. 
Color-magnitude diagrams (CMDs) of the groupings associated with the binaries 
are compared with isochrones, and the majority of the 
groupings are found to have ages $\geq 1$ Gyr, indicating that they 
have an intermediate age. The masses 
of stars at the main sequence turn-off of the groupings are 
estimated, and these provide insights into the initial mass 
of the donor star in each binary system. 
Images from the WISE Allsky survey are also used to search for 
circumsystem envelopes. Extended thermal emission is found around 
six systems in W2 (i.e. $\sim 4.5\mu$m) images.

\end{abstract}

\section{INTRODUCTION}

	\citet{pk1978} identified a group of close binary systems 
(CBSs) that have prominent emission features in the UV. 
In addition to excess levels of UV emission, these 
systems exhibit other observational peculiarities, including differences 
between light curve maxima \citep[e.g.][]{dm1984}, 
and circumsystem envelopes \citep{desetal2015}.
The prototype system is W Serpentis, and a number 
of other candidate systems have been identified \citep[e.g.][]{wiletal1984}. 

	The observational evidence suggests that 
W Ser systems are experiencing rapid mass transfer that occurs after 
the initially more massive star has evolved to fill its Roche lobe. 
The period of a system will decrease during the earliest stages of mass 
transfer, causing elevated mass flow rates until 
well after the mass ratio is reversed. The final result 
of the first episode of such rapid mass transfer may be a 
classical Algol configuration. Modelling of systems in which mass transfer 
is initiated prior to evolution off of the main sequence 
suggests that the initial rate of mass transfer 
may be roughly an order of magnitude higher 
than that in Algol systems \citep[e.g.][]{nelandegg2001}. 
W Ser systems are thus of great importance for 
understanding the evolution of CBSs in general, and Algols in particular.

	The initial masses and ages of the stars in a binary system are of 
fundamental importance for understanding the relationship 
between their initial and final states. The moderately long 
orbital periods of W Ser systems are consistent with the donor star having 
evolved off of the main sequence \citep[e.g.][]{vanetal2011}, and so 
they are likely undergoing Case B mass exchange. Indeed, the spectral types 
of the mass-losing stars in W Ser systems are consistent with a giant 
or supergiant luminosity class and suggest initial masses that 
are well in excess of solar. Still, caution must 
be exercised when drawing even broad conclusions about initial masses 
and evolutionary state as mixing due to rotation and tidal effects will play 
a role in defining evolution within CBSs. These processes will prolong the 
evolution of a star when compared with its static, isolated counterpart, 
while also potentially altering the chemical content of its outer atmosphere 
\citep[e.g.][]{maeandmey2000, sonetal2013}. The 
result of rotation and tidal mixing will make a star appear more 
luminous than a single non-rotating star of the same mass. Models 
of evolution in CBSs demonstrate that the evolution of the initially 
more massive star departs substantially from that expected for an 
isolated star of the same mass \citep[e.g.][]{vanetal2011}. 

	The masses of the component stars in W Ser 
systems are not known with certainty as the spectrum of the star 
that is gaining mass tends to be difficult to detect. These uncertainties 
are demonstrated in Table 1 of \citet{dav2022a}, which shows the sizeable 
range in component masses that have been suggested for V367 Cyg. The 
uncertainty in the component properties of W Ser systems in general is due in 
part to the evolved nature of the donor star and its 
brightness when compared with that from the less evolved receiver. 
Large-scale mass transfer also forms an accretion disk and other light 
emitting features that will contribute spectrophotometric 
features that do not track the orbital motions of 
the stars, while also masking light from the 
receiving star. A circumsystem shell may also form that complicates 
efforts to identify spectroscopic features from either star. 
These factors conspire to make W Ser systems single-line 
spectroscopic binaries.

	Given the challenges described above, observational 
limits on the mass and age of the donor star in W Ser 
systems that are independent of the orbital motions of 
the components are of obvious interest. One approach is to identify 
remnant clusters or moving groups that are associated with these 
systems. It is not unreasonable to expect that companions to at least some 
W Ser systems might be present: if the donor stars have initial masses 
of a few times solar then they will have ages of a few tens to a 
few hundreds of Myr. Clusters in this age range are present in the 
solar neighborhood, although they are admittedly the rump of what was a larger 
cluster population that has since been depleted. 

	In the current study astrometric information is used to search for 
stars that are physically associated with eight CBS systems that 
are either confirmed W Ser systems or related objects. Parallax and 
proper motion measurements that are compiled in the GAIA DR3 \citep[]{gai2016, 
gai2022} are used to identify stars with distances that are 
comparable to those of the W Ser systems and that have similar motions on 
the sky. Photometric measurements made by GAIA are used to 
construct color-magnitude diagrams (CMDs) of the systems. 
Comparisons are then made with model isochrones. 

	Selected observational properties of the target systems 
are listed in Table 1. Magnitudes are in the GAIA system 
\citep{joretal2010}. The targets are eclipsing binaries or ellipsoidal 
variables, and the $G$ magnitudes do not account for variations in 
light levels due to eclipses and/or activity within 
the system. The targets include all but one of 
the systems ($\beta$ Lyrae) discussed by \citet{pk1978}. 
While it is arguably the best-studied W Ser system, $\beta$ Lryae is 
saturated in the GAIA observations and so is not considered here. 
This being said, \citet{bas2019} present evidence of a possible 
cluster associated with $\beta$ Lyrae. Three systems from the list 
compiled by \citet{wiletal1984} with properties that indicate 
that they are W Ser systems or related objects 
are also included. Many of these systems are the subject 
of an on-going spectroscopic monitoring campaign 
at the Dominion Astrophysical Observatory \citep[e.g.][]{dav2022a}. 

\begin{center}
\begin{deluxetable}{cccccccl}
\tabletypesize{\scriptsize}
\rotate
\tablecaption{System Properties}
\tablehead{Name & RA & Dec & $\pi$\tablenotemark{a} & $\mu_\alpha$\tablenotemark{b} & $\mu_\delta$\tablenotemark{b} & G & Source \\
 & (2000) & (2000) & (mas) & (mas/year) & (mas/year) & (magnitudes) & }
\startdata
W Ser & 18:09:51 & --15:33:00 & 1.134 & 0.59 & 3.06 & 8.74 & \citet{pk1978} \\
RX Cas & 03:07:56 & 67:34:39 & 1.818 & --3.11 & --2.11 & 8.42 & \citet{pk1978} \\
SX Cas & 00:10:42 & 54:33:29 & 1.256 & --0.82 & --2.36 & 8.75 & \citet{pk1978} \\
W Cru & 12:11:39 & --58:47:01 & 0.476 & --6.23 & 0.18 & 8.02 & \citet{pk1978} \\
V367 Cyg & 20:48:00 & 39:17:16 & 1.054 & --3.36 & --3.76 & 6.91 & \citet{pk1978} \\
 & & & & & & & \\
V1507 Cyg & 19:48:42 & 29:24:08 & 1.058 & 1.55 & --5.76 & 6.89 & \citet{wiletal1984} \\
V644 Mon & 06:57:09 & --10:49:28 & 0.877 & --0.52 & --0.17 & 6.68 & \citet{wiletal1984} \\
V356 Sgr & 18:47:52 & --20:16:28 & 1.478 & 0.52 & --4.85 & 6.90 & \citet{wiletal1984} \\
\enddata
\tablenotetext{a}{Parallax}
\tablenotetext{b}{Proper motion, with $\alpha$ = Right Ascension, $\delta$ = Declination}
\end{deluxetable}
\end{center}

	All of the systems are within $\pm 10$ degrees of the Galactic Plane, 
and so are in moderately crowded fields. Six of the systems 
are within 1 kpc of the sun, and some have previously been 
associated with clusters or groupings. For example, 
\citet{hal1989} suggested that V644 Mon may be part of Canis Majoris OB1.
W Cru is the most distant system in the sample, and is located in the 
Scutum-Centaurus spiral arm in close projected proximity to areas of active 
star formation. However, the parallax measurements used here indicate that 
neither V644 Mon nor W Cru are associated with active star-forming regions. 
A more detailed discussion of possible membership in known 
clusters and associations is provided in Section 7. 

	Additional clues into the evolution of the systems can be gleaned 
from their circumsystem environments. Simulations suggest that 
dust shells form around systems that are undergoing rapid mass 
transfer, and models suggest that emission from the shells may be detected 
in nearby (i.e. within a few hundred parsecs) systems with 
space-based infrared (IR) observatories that have only a modest aperture 
\citep{desetal2015}. The shells are expected to dissipate after the pace of 
mass transfer declines \citep[e.g][]{nelandegg2001, desetal2015}, and so 
studies of spatially extended dust shells have the 
potential to identify systems that are in the early stages of mass transfer. 
In the current study, images obtained as part of the WISE All-Sky survey 
\citep{wrietal2010} are used to examine the environment 
within a half arcmin radius of these systems and search 
for extended envelopes and very red companions 
that may be missed at the shorter wavelengths sampled by GAIA. Evidence of 
extended mid-IR (MIR) emission is found around 6 of the systems.

\section{DATASETS}

\subsection{GAIA} 

	GAIA DR3 (\citep{gai2022} provides basic 
astrometric information that can be used to 
identify potential members of remnant clusters and moving groups. This database 
also contains photometric measurements and reddening estimates that can be used 
to construct CMDs. The photometry is based on aperture measurements, and so 
source confusion is a potential concern, as the 
W Ser stars are in fields near the Galactic Plane. 
However, the selection criteria that are applied here to search 
for companions restrict the sample to moderately bright objects, the photometry 
of which is less susceptible to contamination than is the case for stars near 
the photometric faint limit. Evidence to support the robustness of the 
photometry comes from the CMDs of star clusters constructed from GAIA photometry
that are discussed by \citet{babetal2018}. These show well-defined 
sequences that extend to magnitudes that are fainter than those considered 
here.

\subsection{WISE}

	The WISE All-Sky survey \citep{wrietal2010} recorded images in four 
bandpasses that span the $3 - 27\mu$m wavelength interval with angular 
resolutions $\sim 6 - 12$ arcsec ($\sim 3000 - 6000$ AU at a distance of 
1 kpc). The processed results are available on line (Wright et al. 2019). 
Here we focus on images taken through the W2 ($\lambda_{cen} = 
4.6\mu$m) filter. Processed image tiles were downloaded from the 
IPAC website \footnote[1]{https://irsa.ipac.caltech.edu/applications/wise/}, 
and these were used to search for extended circumsystem emission. 
Evidence for resolved emission around some of the systems is presented in 
Section 6.

\section{SOURCE SELECTION}

	Potential companions of the W Ser systems are identified using 
a combination of the parallaxes, proper motions, and locations on the sky 
of sources that are compiled in the GAIA DR3. It is anticipated that any 
companions will belong to diffuse clusters or moving groups. 
Two samples are generated with this expectation in mind, and the selection 
criteria for each of these samples are described in the following sub-sections. 

\subsection{The Baseline Sample}

	An initial list of possible companions was made by 
identifying stars that have projected distances 
on the sky within $\pm 10$ pc of each system, where the angular 
extraction radius is calculated using the GAIA DR3 parallax measurement. 
A 10 pc radius samples an area that is $4\times$ larger than 
that subtended by the main body of the Hyades. While generous, such a large 
initial search area was employed as heretofore unidentified old 
clusters and moving groups are expected to be diffuse.

	Stars in the search area were then filtered according to parallax. 
While seemingly a straight-forward process, 
potential systematic uncertainties in the parallaxes 
should be considered when selecting a suitable range to search for 
companions. \citet{linetal2021} identify 
brightness, color, and location on the sky as possible sources of systematic 
errors in GAIA parallaxes. These uncertainties have been reduced with 
progressive GAIA data releases \citep[e.g.][]{staandtor2021,renetal2021}. 

	Location on the sky is likely not a concern 
for the selection of possible companions in the context of systematic 
uncertainties in the parallaxes. This is because the search is restricted to a 
comparatively small angular extent (less than a 
degree in radius) centered on each system. All parallaxes 
for candidate companions to a W Ser system should then have a 
similar location-based bias. While such a bias may affect the measurement 
of absolute quantities such as distance, absolute magnitude, and age, 
the selection of a sample of possible companions to the W Ser 
systems will not be affected.

	Systematic uncertainties that are related to the color and 
brightness of a source are potentially of greater concern, as 
\citet{linetal2021} find that the parallax of sources with the same distance 
may differ by a few tens of mas depending on their brightness and color. 
However, there are hints that such biases may not be significant for the 
sources examined in this study. The faint limit imposed by selection 
criteria that are described below is in the range $G = 16 - 18$.
\citet{muzetal2019} find no systematic trend in parallax among blue stars 
in the young open cluster NGC 2244 down to $G \sim 16$ 
in Figures 4 and 5 of their paper. As for systematic errors in parallax 
that might arise due to color, in most cases the final samples of 
objects found in the current work are dominated by blue stars, reducing 
the influence of color-related systematics. The CMDs of 
distant open clusters discussed by \citet{babetal2018} 
provide further evidence that systematic errors in parallax due 
to brightness and color may not be a major concern.

	Even though there is evidence that systematic errors in brightness and 
color may be small, we take the cautious approach of applying a wide parallax 
selection filter. The maximum systematic errors in parallax due to color are 
comparable to the uncertainties in the parallaxes of the W Ser stars at the 
1 to $2\sigma$ levels. To account for possible systematic 
errors of this size, stars were extracted from 
the database that had parallaxes within $2\sigma$ 
of the parallax of the W Ser star. This sampled line of sight depths 
ranging from $\pm 15$ pc for RX Cas (the nearest system) 
to $\pm 270$ pc for W Cru (the furthest system). 
While fostering contamination from objects that are 
not related to the W Ser star, this criterion allowed objects 
that may otherwise be missed due to a systematic error in parallax 
but could still be possible members of a cluster/moving group to be included. 
Stars that are not related to the W Ser systems are identified 
using proper motion measurements (see next section).

	The sample was further restricted to sources 
that have $1\sigma$ random uncertainties in parallax 
that are less than the $2\sigma$ random error estimates for the 
host W Ser systems. Not only does this limit the number of 
interlopers with uncertain parallaxes, but it also effectively 
sets the faint limit of the samples to $G \sim 16 - 17$. This is a relatively 
benign filter in the context of estimating system ages, as stars with large 
uncertainties in the parallax tend to be significantly fainter 
than the main sequence turn-off (MSTO) in intermediate age groups at the 
distances examined here, and so contain little information about age. 

	Filtering based on angular extent on the sky and distance 
results in a list of candidate companions that 
are refered to as the baseline sample. 
GAIA DR3 lists proper motions in declination and right 
ascension, and the proper motion diagrams 
for baseline sample objects are shown in Figures 1a and 1b. 
The distribution of points in the proper motion 
diagram of W Cru in Figure 1b is much more compact than for the other 
systems. This is a distance effect -- proper motions will decrease with 
increasing distance among sources that have a given space velocity.

\begin{figure}
\figurenum{1a}
\plotone{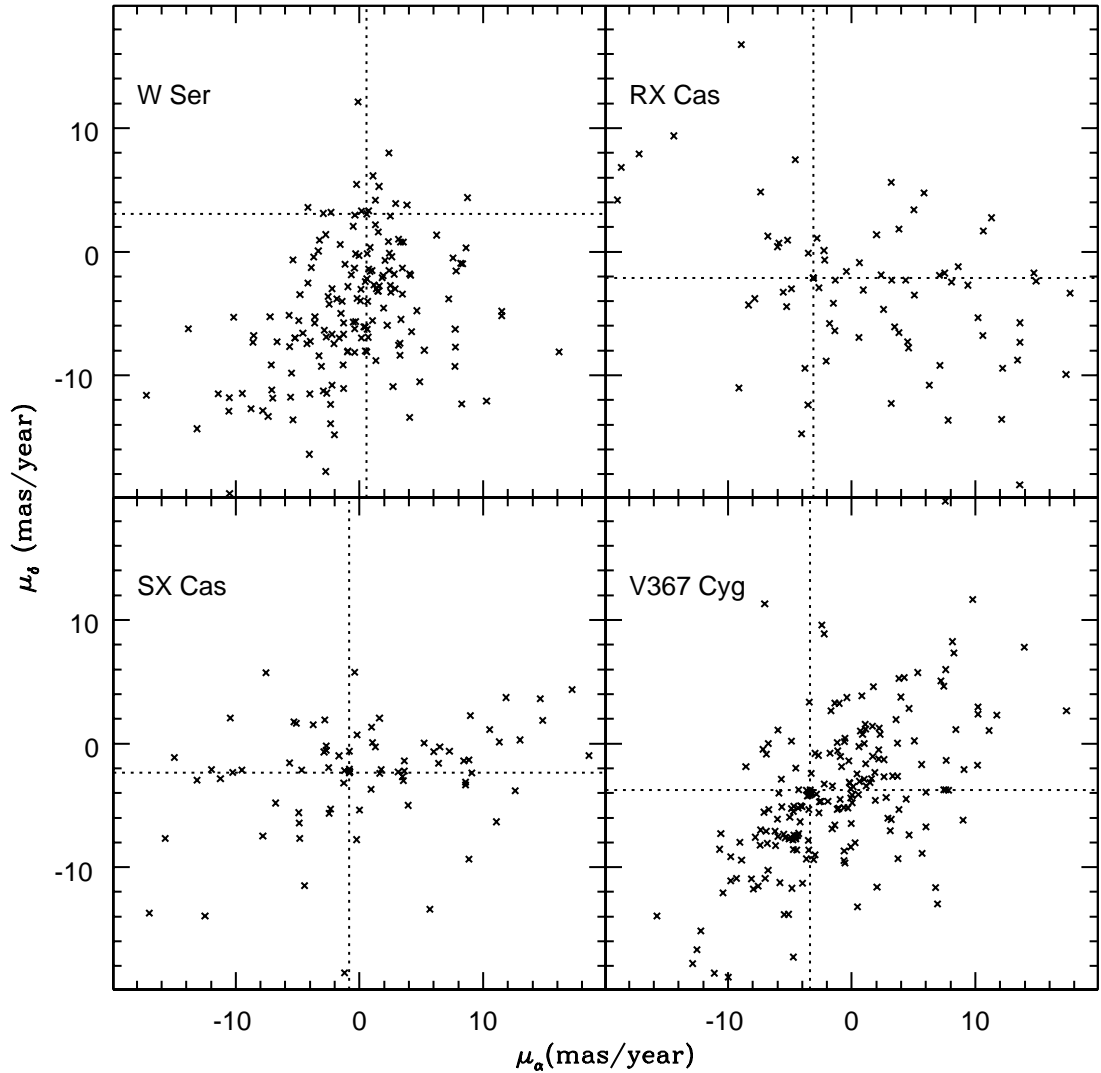}
\caption{Proper motion diagrams of stars in the baseline samples of W Ser, RX 
Cas, SX Cas, and V367 Cyg. $\mu_{\delta}$ and $\mu_{\alpha}$ are proper motions 
in declination and right ascension in mas/year. 
The dotted lines mark the proper motions of the W Ser systems. Typical 
$1\sigma$ random errors in the proper motions of all stars in these 
diagrams are $\pm 0.02 - 0.03$ mas/year along each axis. 
The vast majority of objects in the baseline 
samples are scattered throughout the diagrams, and are 
likely not related to the W Ser systems. Still, there are 
objects that have proper motions that are similar to those of the W Ser systems.
The locations of W Ser and RX Cas indicate that their kinematic properties 
differ in a systematic way from those of most other stars in 
their baseline samples.}
\end{figure}

\begin{figure}
\figurenum{1b}
\plotone{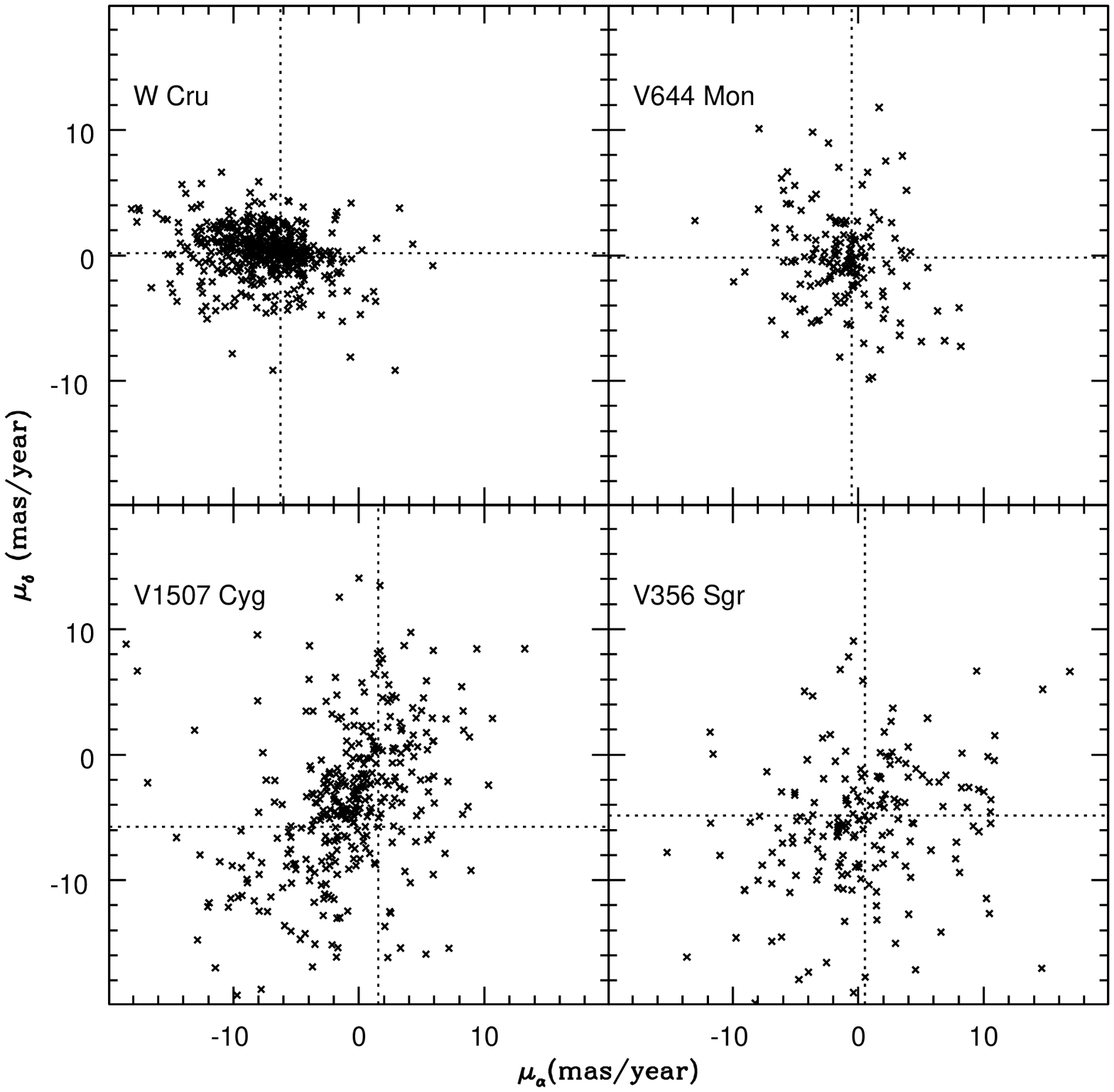}
\caption{Same as Figure 1a, but showing the proper motion diagrams 
of baseline sample stars near W Cru, V644 Mon, V1507 Cyg, and V356 Sgr. 
There are obvious concentrations on the proper motion plane around 
V644 Mon and W Cru. The large baseline sample for W Cru is due 
to the substantial line of sight distance that is sampled for this 
system. A concentration of objects is evident in the proper motion 
plane of V1507 Cyg that is offset from the proper motions of that system.} 
\end{figure}

	Evidence for clumping is seen in these panels, and in some cases these 
concentrations are not centered on the proper motions of the W Ser systems, 
which are indicated with the dotted lines in these figures. 
A prime example is V1507 Cyg, which is offset from a denser clump of sources 
on the proper motion diagram. The presence of proper motion-based groupings 
that appear to be unrelated to the W Ser stars is perhaps 
not surprising given the low Galactic latitudes that are examined. 

	There is also a tendency for some of the W Ser systems to be offset 
from the centers of the data clouds in the proper motion diagrams. 
W Ser and RX Cas are the most obvious examples, with their 
kinematic properties differing from those of many of their immediate neighbors. 
These systems are not completely isolated on the proper motion plane, 
and it is likely that stars that are near them on the proper motion plane 
share a common kinematic (and hence probably an evolutionary) heritage. The 
identification of these objects is discussed in the next sub-section.
Given that the dynamical properties of a star will evolve with time, 
then one {\it possible} explanation for the locations of some of the W Ser 
systems is that they have ages that differ from the majority of their 
neighbors. 

\subsection{The Final Sample}

	Physical proximity is but one consideration for 
the identification of physically related stellar groupings, 
especially in the low Galactic latitude fields that are examined here. 
Additional filtering was done using projected spatial motions, 
with stars in the baseline sample identified that have 
proper motions that are similar to those of the W Ser stars. 
Selection based on proper motion samples only two axes of a three-dimensional 
vector. Nevertheless, while there are radial velocity estimates for 
some stars, these are not available for the vast majority of stars in 
the baseline samples, and so radial velocities are not considered. 

	The apparent lack of grouping around many of the W Ser systems 
in the proper motion diagrams suggests that in the 
majority of cases any remnant clusters -- if present -- 
are not well populated. W Cru and V644 Mon appear 
to be exceptions, although these are also the most distant 
systems and so will have the most compact 
proper motion distributions. Still, given that many of the 
systems are at similar distances then it is possible to 
determine if there is at least a general tendency for 
these systems to be accompanied by companions. To do this, 
the proper motion diagrams of W Ser, RX Cas, SX Cas, V367 Cyg, and V356 Sgr 
were combined, and the results are shown in the top panel of Figure 2. 
To combine the samples, the proper motions of the host systems were subtracted 
from the proper motions of their baseline samples. The proper motions 
used to construct Figure 2 are then offsets from that of the host system. 
V1507 Cyg was not included due to the concentration of objects 
in its proper motion diagram to the upper left of the W Ser star in 
Figure 1b, as such a collection of stars will skew the comparison in Figure 2.

\begin{figure}
\figurenum{2}
\epsscale{0.8}
\plotone{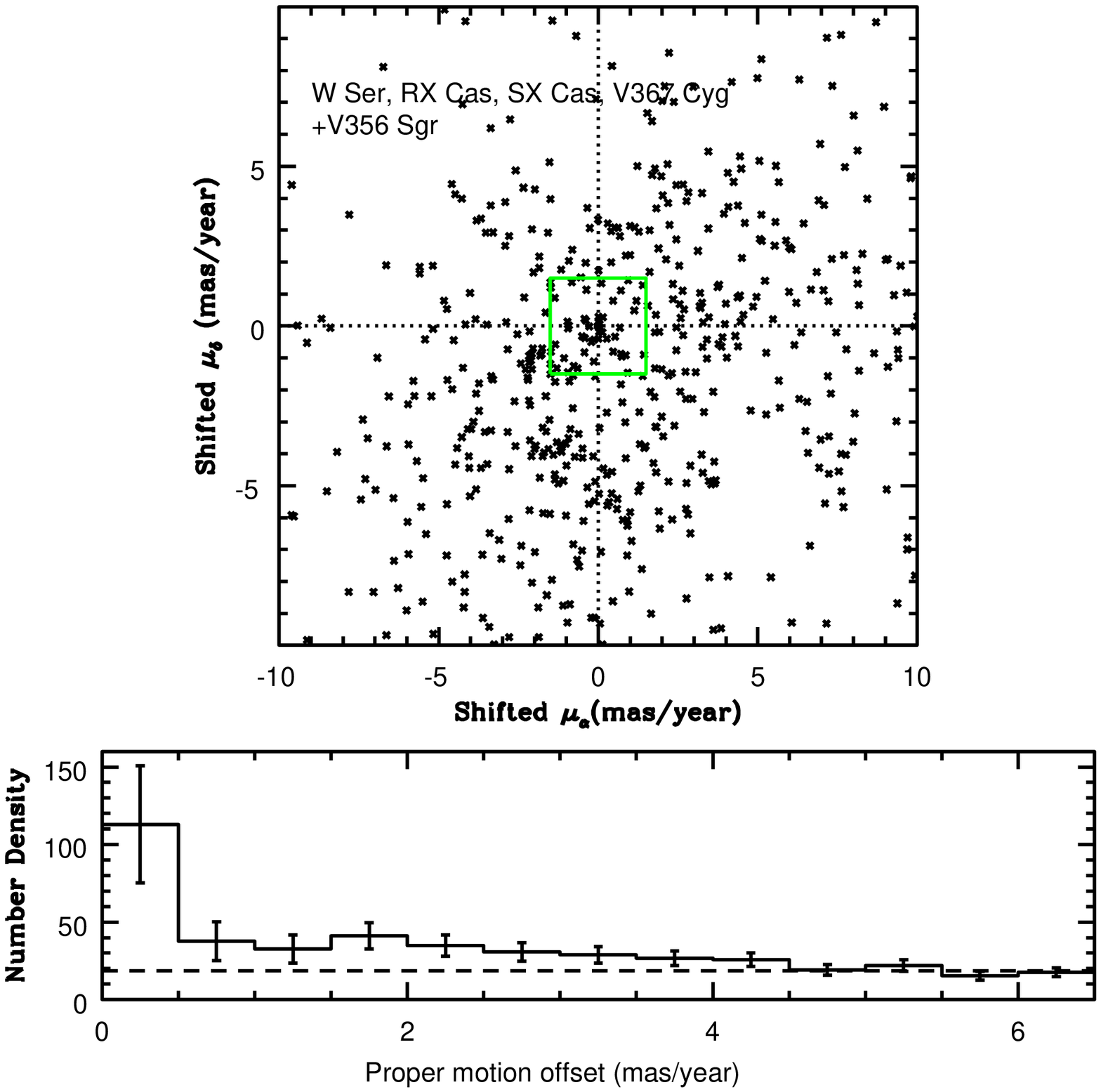}
\caption{(Top panel) Composite proper motion diagram of the 
W Ser, RX Cas, SX Cas, V367 Cyg, and V356 Sgr baseline samples. The 
proper motions of each system have been subtracted 
from those of individual stars to align the samples. 
The origin is at the intersection of the dotted lines. There is 
an apparent concentration of objects at the origin, indicating 
a general tendency for the W Ser systems to have companions. 
The green square is 3 mas/year on a side, and shows the approximate extraction 
area adopted to identify companions for these stars.
(Bottom Panel) Number density of sources in the top panel (i.e. the number 
of objects per mas$^2$/year$^2$) versus the radial offset from the origin. 
The dashed line shows the average density in the four bins with the greatest 
offsets from the origin, where the density in proper motion space appears 
to be stable. While the error bars are large, there is a tendency for the 
density of objects in proper motion space to decrease with increasing 
offset out to $\sim 4.5$ mas/year.}
\end{figure}

	A concentration of objects is evident near the origin in the 
top panel of Figure 2, indicating a general tendency for the W Ser stars to 
have companions that are distributed over spatial scales of 
many parsecs and are moving with them on the sky. This is 
quantified in the lower panel of Figure 2, where the 
density of objects in the proper motion diagram is shown as a function of the 
offset in mas/year from the origin. The number density in the innermost 
bin does not include the W Ser systems, and the error bars reflect counting 
statistics. 

	A uniform distribution of objects on the proper motion plane 
will produce a flat trend in the the lower panel of Figure 2. Not 
including the W Ser systems, there is an excess number of objects that 
have proper motions within 0.5 mas/year of the origin. The 
density in the proper motion plane drops at larger offsets, such that 
the density gradually decreases with increasing offset out 
to 4.5 mas/year. The modest uptick in the number of sources 
near the 2 mas/year offset is due to the clump of objects to the 
lower left of the origin in the upper panel of Figure 2, centered near 
$\Delta\mu_{\delta} \sim -1$ mas/year and $\Delta\mu_{\alpha}\sim 
-2$ mas/year.

	Based on the comparisons in Figure 2, 
potential companions were extracted from the baseline samples by applying 
a $\pm 1.5$ marcsec/year cut around the proper motion of each W Ser system. 
This extraction criterion was adopted as a compromise between the 
desire to sample as many potential companions as possible and the need to 
maintain some degree of contrast with respect to the background sample. 
The one exception is W Cru: as this system is roughly twice as far as the 
other systems then an extraction criterion of $\pm 0.75$ marcsec/year 
(i.e. one half that applied to the other systems) was applied.

	We emphasize that the companions identified in 
proper motion space are not physically 
close to the W Ser systems at the present day. Rather, the 
spatial distributions of these objects on the sky (see Section 4) 
indicate that they are distributed over a large portion of the 
area that is surveyed around the host systems. There is thus no evidence 
that these objects were once close companions to the W Ser systems 
and so may have played a direct role in shaping their evolution. 
Still, the presence of companions suggests that 
the initial properties of the clusters in which the W Ser systems 
formed were such that the clusters have not been completely 
obliterated over $\sim 1$ Gyr timescales.

	Objects that are extracted from the baseline 
samples based on their proper motions form 
the final samples. The fraction of stars in the baseline sample that makes 
it into the final sample varies from system-to-system, although it is 
apparent from Figures 1a and 1b that this fraction is 
small in most cases. SX Cas and RX Cas have the smallest fraction of 
baseline sample objects that make it into the final 
sample, reflecting the small number of objects 
near the intersections of the dashed lines in Figure 1a.

\section{THE PROJECTED DISTRIBUTION AND PHOTOMETRIC PROPERTIES OF POTENTIAL CLUSTER MEMBERS}

\subsection{On-Sky Distribution}

	The projected location of objects on the sky provides insights into 
membership in a stellar group. Nevertheless, while evidence for clustering on 
the sky is obviously desireable when searching for companions, it is 
not essential. In fact, given the low Galactic latitudes of these fields, 
then it might be anticipated that diffuse clusters and moving groups will not 
stand out in a statistical sense when compared with background populations.

	The projected distributions of objects in the baseline and final 
samples are compared in Figures 3a and 3b. At first glance, the 
distributions of the two samples appear to differ in some cases. 
However, what is the statistical significance of such differences? 
The azimuthal and radial distributions of stars in the baseline and 
final samples around each system have been examined to answer this 
question.

\begin{figure}
\figurenum{3a}
\plotone{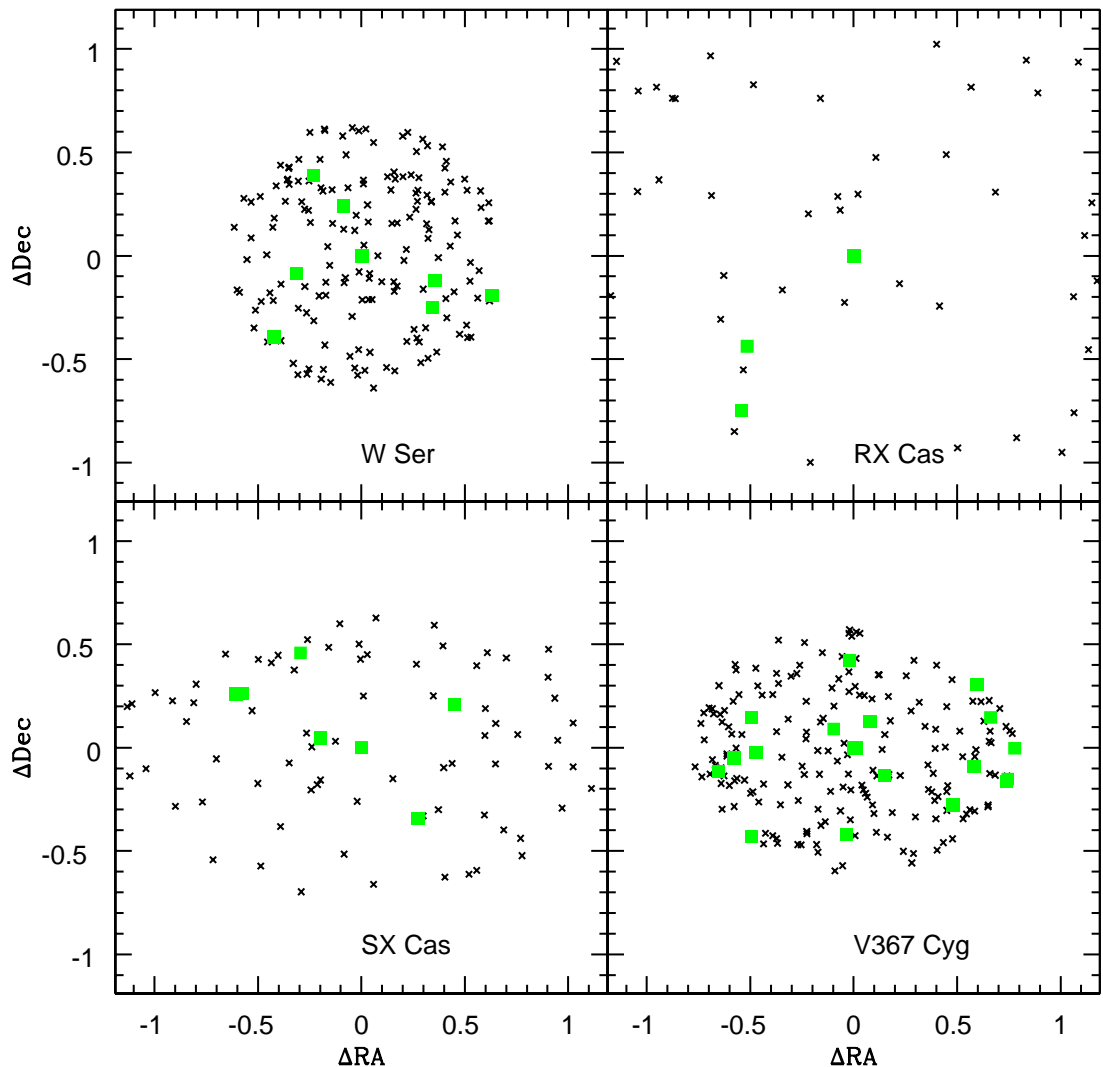}
\caption{Locations on the sky of stars in the baseline (black 
crosses) and final (green squares) samples of W Ser, RX Cas, SX Cas, and 
V367 Cyg. Offsets in right ascension and declination from each W Ser 
system are plotted.}
\end{figure}

\begin{figure}
\figurenum{3b}
\plotone{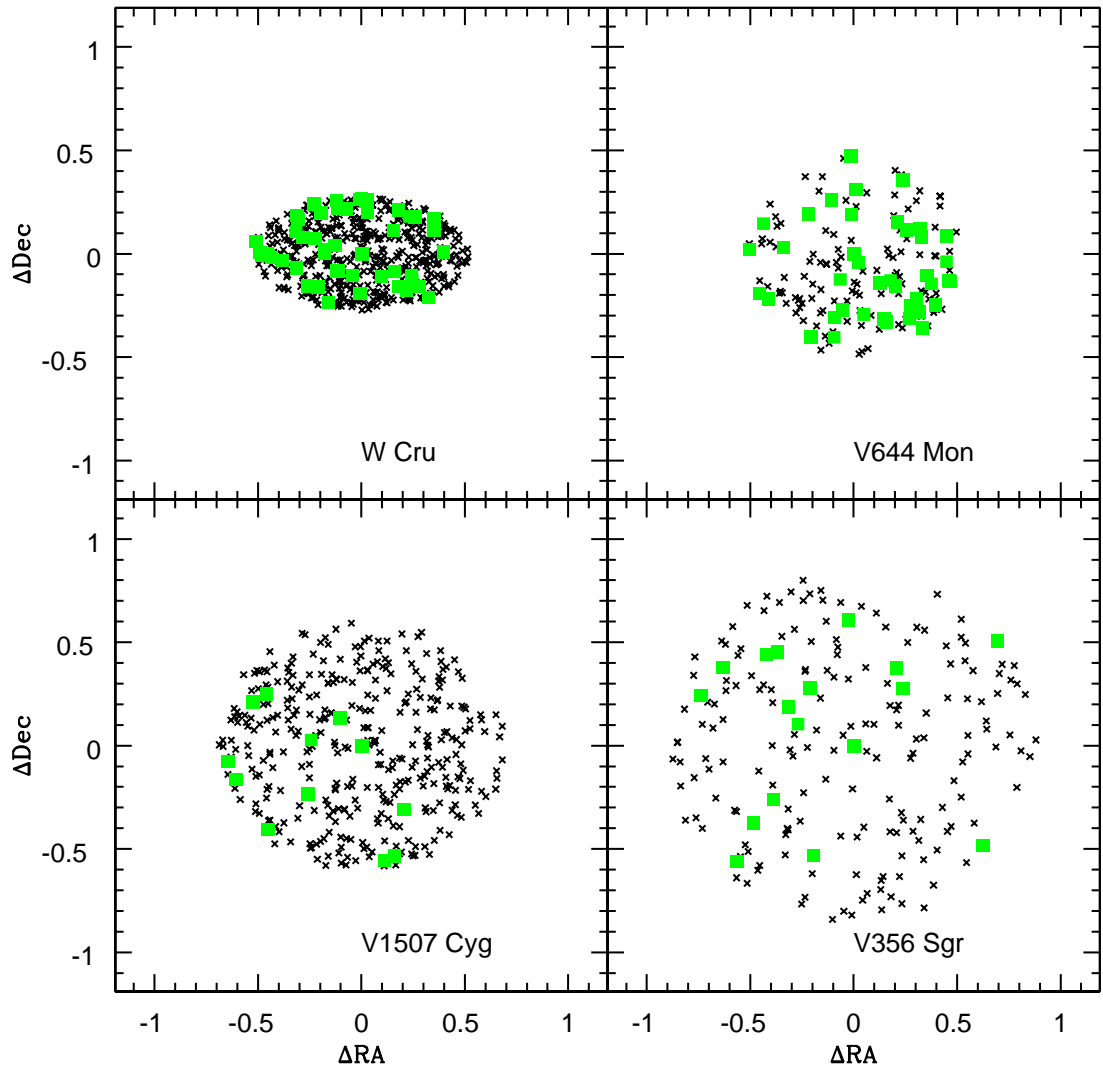}
\caption{Same as Figure 3a, but showing the locations of 
stars near W Cru, V1507 Cyg, V644 Mon, and V356 Sgr. 
The large number of objects in the W Cru final sample is a consequence 
of the large volume sampled around this system.}
\end{figure}

	The angular distributions of the baseline and final samples 
are compared in Figure 4. Position angles were measured about 
each W Ser system with 0 degrees to the East. Cumulative number counts, 
which are the total number of objects up to a given angle 
in 30 degree wide bins, are shown. The results have been normalized to 
the number of objects in each sample. 

\begin{figure}
\figurenum{4}
\plotone{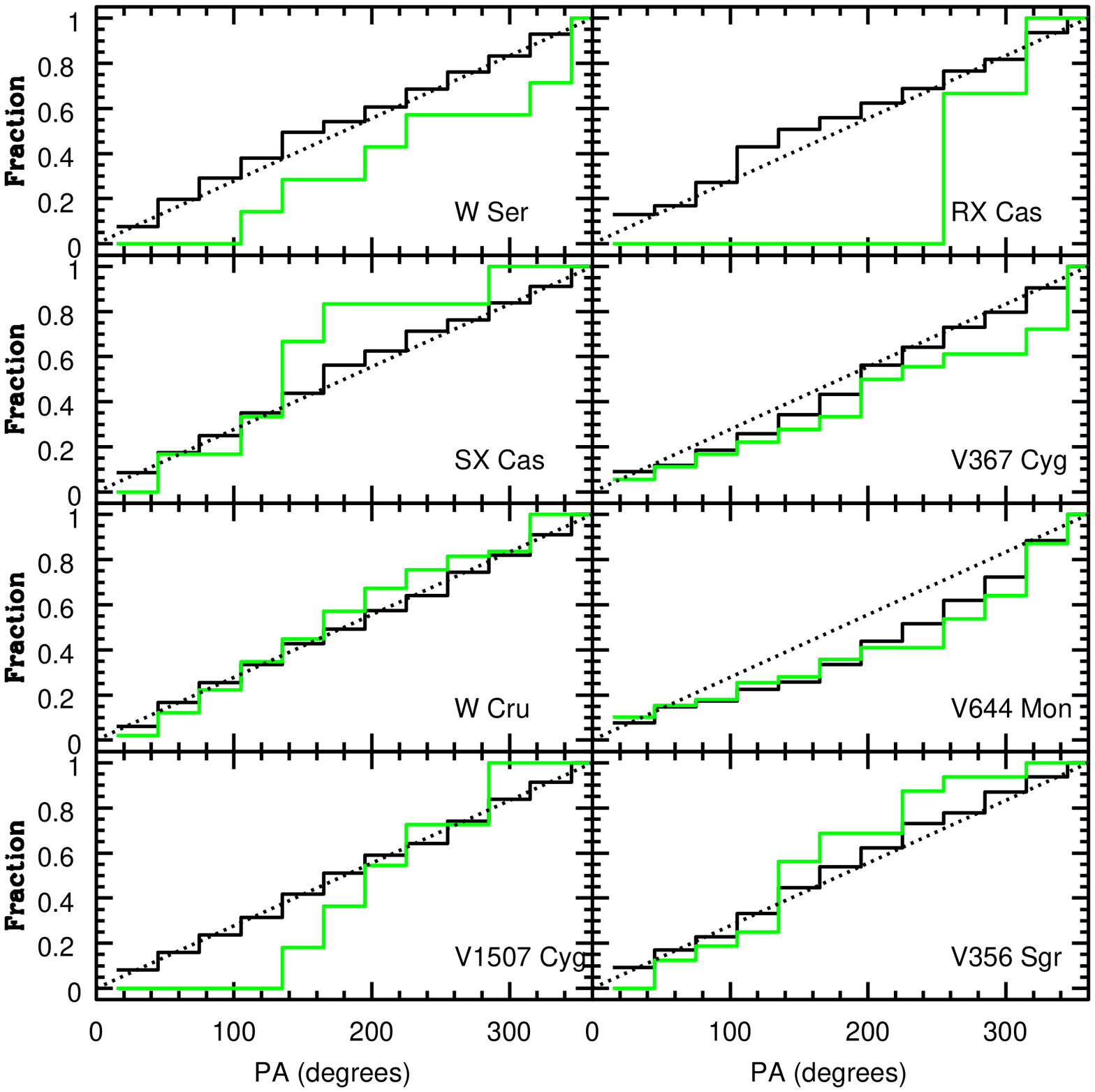}
\caption{Angular distributions of stars in the baseline 
(black line) and final (green line) samples. 
Cumulative number counts in 30 degree wide bins are shown along the y axis. 
Position angles (PAs) were measured centered on the W Ser systems, with 
0 degrees to the East. The distributions have been normalized 
to the total number of objects in each sample. Objects that are 
uniformly distributed on the sky define a diagonal that is 
indicated by the dotted lines. In most cases the baseline samples 
are consistent with a uniform distribution. Possible exceptions are V367 Cyg 
and V644 Mon, where there is a tendency for objects to be concentrated 
between 260 and 360 degrees. A Kolmogorov Smirnov test indicates that the 
angular distributions defined by the baseline and final samples 
of RX Cas and V1507 Cyg differ at the 75 -- 85\% confidence level.}
\end{figure}

	A sample of objects that is uniformly distributed on the sky will form 
a diagonal sequence in Figure 4, and the dotted lines in each panel mark this 
trend. The baseline samples for the majority of systems more-or-less follow 
the trend expected for a uniformly distributed sample, although V367 Cyg 
and V644 Mon are possible exceptions. There is a tendency in both cases 
for objects in the baseline sample to concentrate between 180 and 360 
degrees, indicating a lop-sided distribution on the sky. A Kolmogorov-Smirnov 
(KS) test confirms that these differences are significant in excess of the 
90\% confidence level. 

	V367 Cyg and V644 Mon have locations on the sky that are close to 
star-forming regions that have high stellar densities. However, the lop-sided 
angular distributions in the baseline samples of these systems are likely not 
due to nearby young clusters. V367 Cyg is roughly a degree away from the 
center of the open cluster M29, and the parallaxes of V367 Cyg 
and M29 differ by an amount that far exceeds the range used 
to define the baseline sample. As for V644 Mon, the location of that system on 
the sky coincides with Group 12 identified by \citet{sanetal2021} 
in their study of CMa OB1. While some of the stars in this group 
have proper motions that overlap with V644 Mon, the range 
of parallaxes that are used to extract the V644 Mon baseline sample 
do not overlap with those of objects in that 
group \citep[Figure 2 of][]{sanetal2021}. The proper motion 
diagrams of V367 Cyg and V644 Mon also do not show noticeable 
concentrations that are offset from the proper motions of the systems, 
as might be expected if there was substantial 
contamination from a kinematically unrelated cluster. In fact, V644 Mon is near 
the center of the dominant cloud of objects in the 
proper motion diagram. We suspect that the asymmetric 
distributions of baseline objects near V367 Cyg and V644 Mon are due to 
foreground obscuration.

	If the baseline samples are dominated by field stars then 
a difference between the angular distributions of the baseline and 
final samples might be expected if the W Ser systems are in fossil groups. 
For example, a lop-sided distribution in the final 
sample could occur if a W Ser system is embedded in a cluster or 
moving group but is not near its geometric center. A skewed angular 
distribution might also occur if the W Ser systems formed in 
filaments that have not yet dissipated, such has been suggested 
for the massive binary V640 Mon \citep{dav2022b}.

	The position angle distributions of the baseline and final samples in 
Figure 4 have been compared with the KS statistic. The most statistically 
significant differences between the angular distributions 
of these two samples are those for RX Cas (85\% confidence) and 
V1507 Cyg (75\% confidence). V1507 Cyg is located on the sky midway between 
the open clusters NGC 6834 and Turner 9. However, contamination from 
stars in those clusters do not cause the skew in the NGC 1507 final 
sample distribution, as both clusters have parallaxes that are 
far smaller than that of V1507 Cyg \citep[]{canandand2020}, and so 
do not contribute stars to the baseline and final samples.

	The radial distributions of sources centered on the W Ser systems is 
another probe of grouping if it is assumed that the W Ser system is in a 
cluster and if the baseline stars are an unclustered field component. The 
radial distributions of the final and baseline samples are compared in 
Figure 5. As with Figure 4, the panels in Figure 5 show cumulative 
number counts up to a given angular offset from each W Ser system. 
A bin width of 0.05 degrees has been adopted. System-to-system 
differences in the total angular coverage on the sky reflect the 
range in system distances. 

\begin{figure}
\figurenum{5}
\plotone{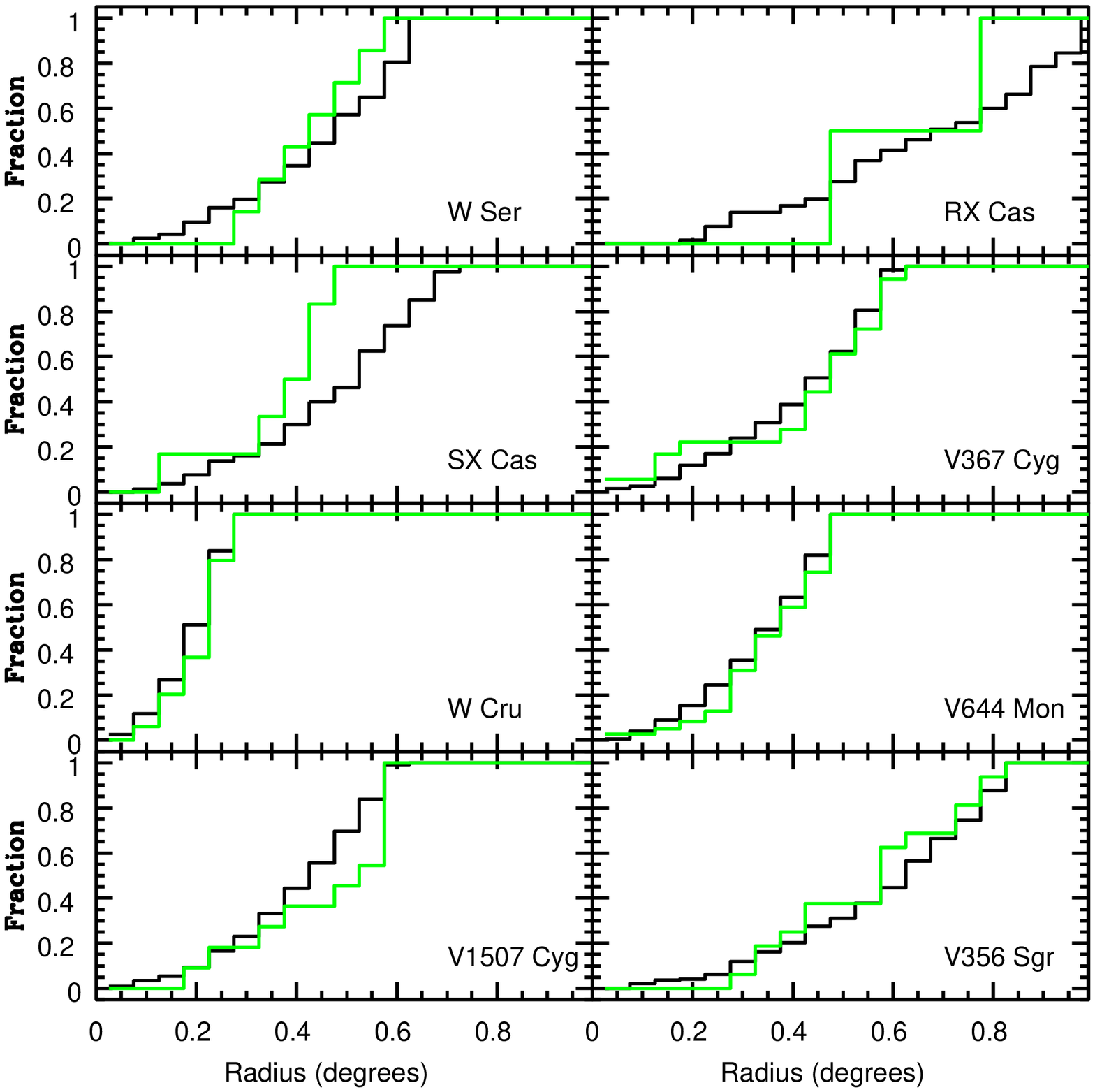}
\caption{Same as Figure 4, but showing radial distributions centered 
on the W Ser systems. Cumulative number counts have been measured in 
0.05 degree angular bins on the sky centered on the W Ser systems, 
with the distributions normalized to the total 
number of counts in each sample. System-to-system differences 
in the slopes and total angular extents of the baseline distributions 
are due to the wide range in system distances. With the exception of SX Cas, 
and to a lesser extent W Cru and V1507 Cyg, the radial distributions 
of the baseline and final samples are not significantly different.}
\end{figure}

	The KS statistic indicates that the baseline and 
final radial distributions of SX Cas differ at the 
95\% confidence level. This difference is in the sense that stars 
in the final sample of SX Cas have a more compact distribution on the sky than 
those in the baseline sample. That the angular distributions of the 
baseline and final samples of SX Cas do not differ suggests that SX Cas 
is likely close to the center of a structurally distinct 
stellar ensemble. Weaker evidence for clustering is found 
for W Cru and V1507 Cyg, where the radial distributions of the 
final and baseline samples differ at roughly the 70\% confidence level. 

\subsection{CMDs}

	The CMDs of the baseline and final samples 
are shown in Figures 6a and 6b. The faint limits 
fall between $G \sim 16$ and 17, and these are 
largely defined by the maximum allowable error in the parallaxes. 
The primary region of interest in the 
CMDs when estimating ages is the upper portions of the main sequence, 
and this is well above the faint limit in all cases.
The main sequence in most cases tends to be well-defined, as expected 
given the moderately small dispersion in distance that results from selection 
based on parallax, coupled with the expectation of solar-like 
metallicities based on the locations of the target fields at low Galactic 
latitudes. The main sequences in the baseline sample CMDs 
of W Ser, SX Cas, V644 Mon, and V356 Sgr have a width 
along the color axis that is comparable to the dispersion 
in the CMDs of clusters examined by \citet{babetal2018}.

\begin{figure}
\figurenum{6a}
\plotone{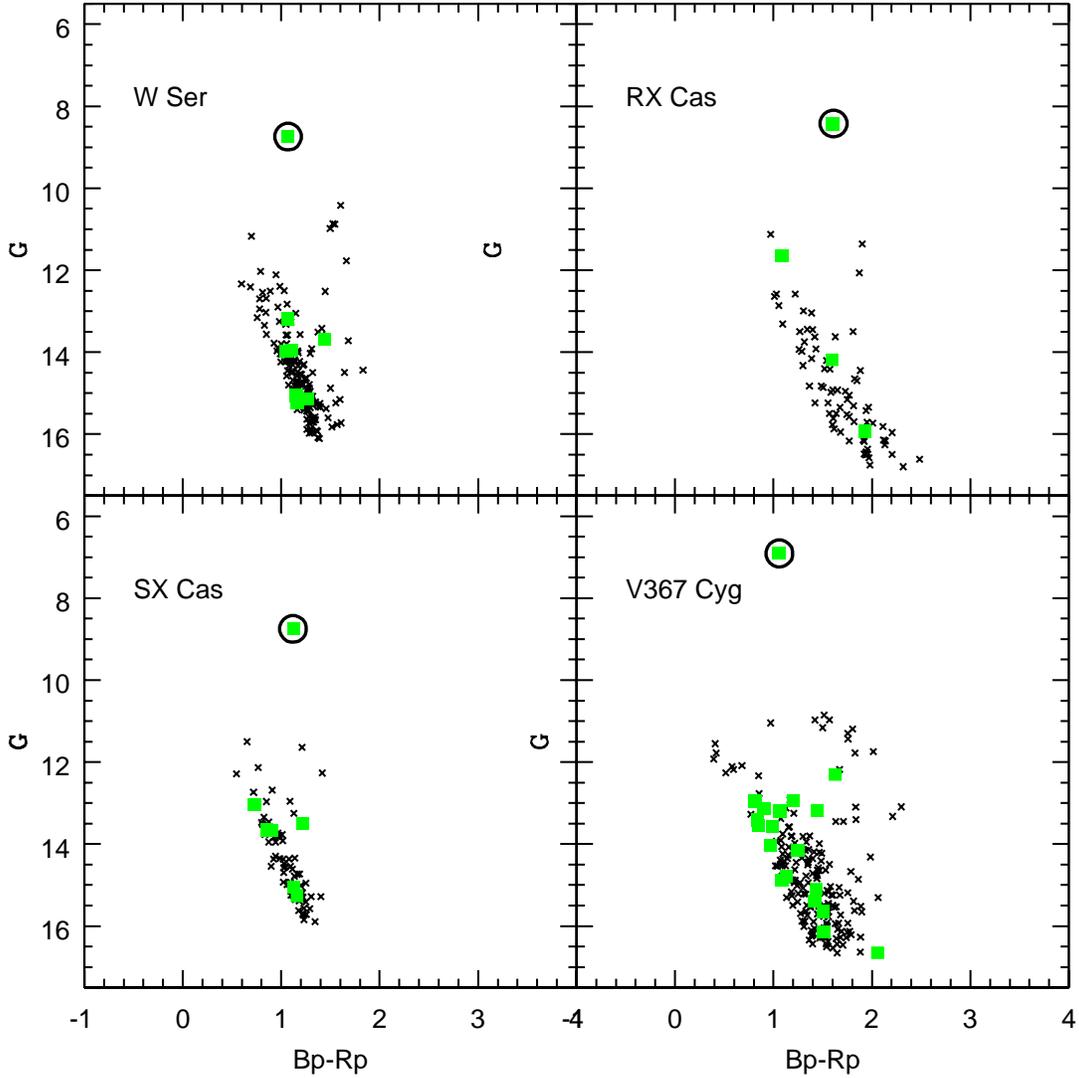}
\caption{$(G, Bp-Rp)$ CMDs of stars in the baseline (black crosses) and 
final (green squares) samples of W Ser, RX Cas, SX Cas, and V367 Cyg. 
The W Ser system is the brightest object in each CMD, and these are 
circled in black. The main sequence is the doninant feature in each CMD. 
The stars with red colors that are above and to the 
right of the main sequence are likely SGB stars that belong to 
an intermediate age field population.} 
\end{figure}

\begin{figure}
\figurenum{6b}
\plotone{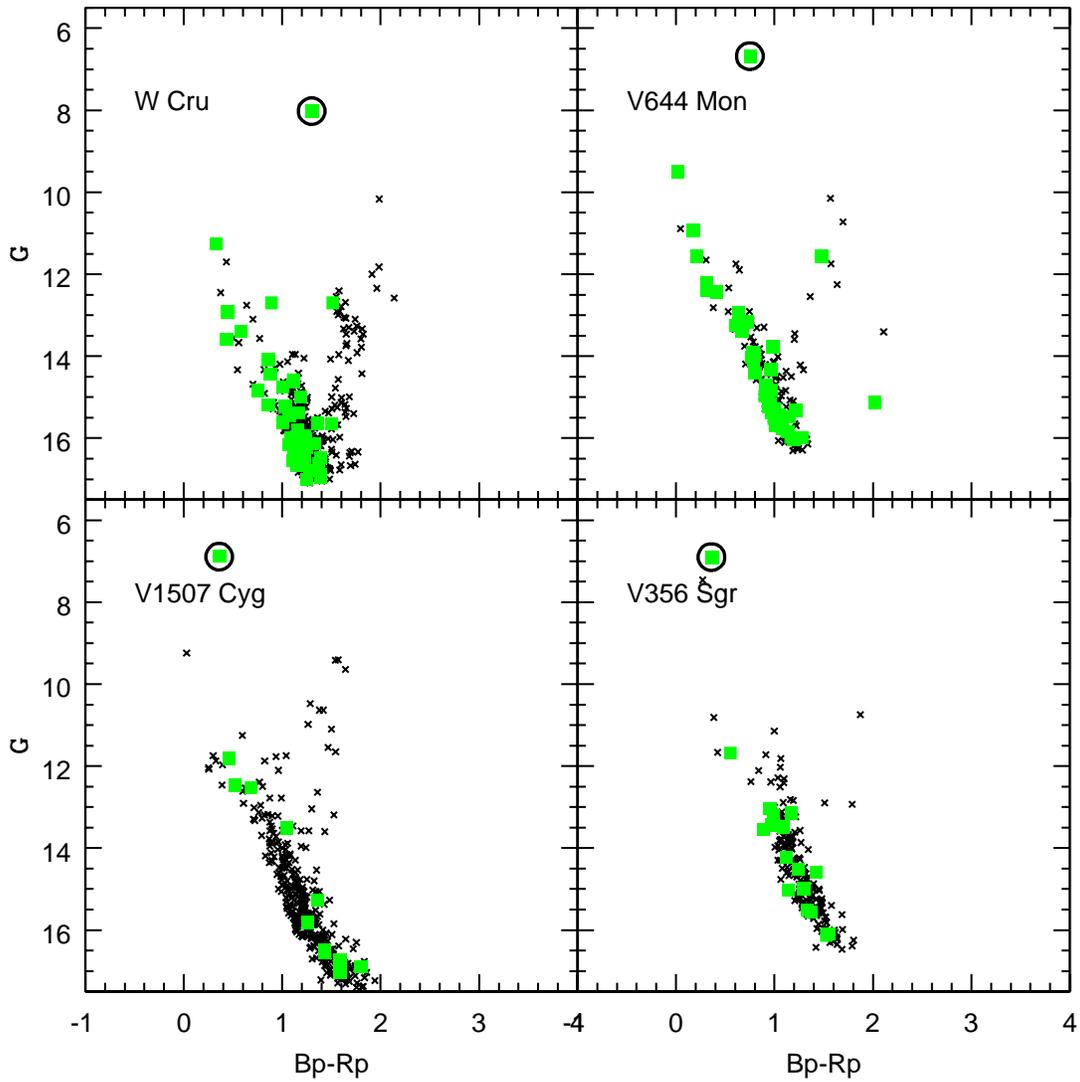}
\caption{Same as Figure 6a, but showing the CMDs for W Cru, V644 Mon, V1507 
Cyg, and V356 Sgr. SGB sequences are seen above and to the 
right of the main sequence in the baseline sample CMDs of W Cru, 
V644 Mon, and V1507 Cyg.}
\end{figure}

	The narrow main sequence notwithstanding, there is evidence that the 
baseline samples contain stars with a range of ages, as expected if they 
contain a dominant field component. A red 
sub-giant branch (SGB) with a $G$ brightness that is comparable to, 
or fainter than, the apparent main sequence turn-off (MSTO) 
of the baseline sample is present in most CMDs. This is a standard feature in 
disk field CMDs \citep[e.g.][]{babetal2018}, and is due to the presence of 
stars that have older ages than those near the MSTO. There is also scatter 
among baseline sample stars in the CMDs near the MSTO. 
While an age dispersion is one possible explanation, 
binarity, differences in rotation rates, and differential 
reddening can also broaden this part of the CMD.

	Unlike the baseline samples, the final samples might be expected to 
consist of a homogeneous ensemble of objects, given that they have been 
selected with the additional criterion of similar kinematic properties. 
The brightest main sequence stars in the final samples of W Ser, SX Cas, 
and V367 Cyg tend to be fainter than main sequence objects 
in the baseline samples, indicating that these systems 
may not be as young as neighboring stars. This is not a surprising result 
given that the systems are at low Galactic latitudes, where a 
population of young field stars might be expected. In contrast, the brightest 
blue stars in the final sample CMDs of W Cru, V644 Mon, and 
V1507 Cyg are at or near the MSTO of the baseline samples associated with 
those systems.

\section{AGE AND MASS ESTIMATES}

\subsection{Comparisons with Isochrones}

	Comparisons have been made with the 
MESA Isochrones and Stellar Tracks \citep[MIST:][]{paxetal2011,
paxetal2013,paxetal2015,choetal2016,dotetal2016}. As the W Ser 
systems are in the Galactic disk within a few kpc of the Sun 
then a solar metallicity is assumed. The sequences 
used here assume non-rotating stars (i.e. $\frac{v}{v_{crit}} = 0$), 
and adopting models with $\frac{v}{v_{crit}} = 0.4$ will change the 
positions of the isochrones by $\sim 0.1$ mag towards brighter values 
along the vertical axis for the range of ages found for the stellar groupings.
The use of these rotating models will then not have a major impact on age 
estimates. The GAIA database includes reddenings and selective extinction 
estimates for the vast majority of stars in the baseline sample, and the 
isochrones have been shifted to account for interstellar extinction 
by applying the median A$_G$ and A$_{BpRp}$ values 
for the stars in the baseline samples. 

	The CMDs are compared with isochrones in Figures 7a and 7b. 
The isochrone placement for each system is based on the parallax 
measurement of that system with no correction for potential systematic 
effects due to brightness, color, etc. The isochrones 
have also been shifted using the median A$_{G}$ and A$_{GpRp}$ values of 
objects in the baseline samples that are listed in the GAIA database. 

	Age estimates are listed under the system names in these figures. These 
are based on a 'by eye' matching of isochrones with 
the final sample data. These estimates are anchored on 
the MSTO and, when available, the sub-giant branch (SGB). 
With the exceptions of W Cru and V644 Mon, the CMDs of stars in the 
final sample are matched with isochrones that have ages log(t$_{yr}$) $\geq 9$. 
The comparisons with isochrones thus suggest that the classical 
W Ser systems identified by \citet{pk1978} are intermediate age objects. 

\begin{figure}
\figurenum{7a}
\epsscale{0.9}
\plotone{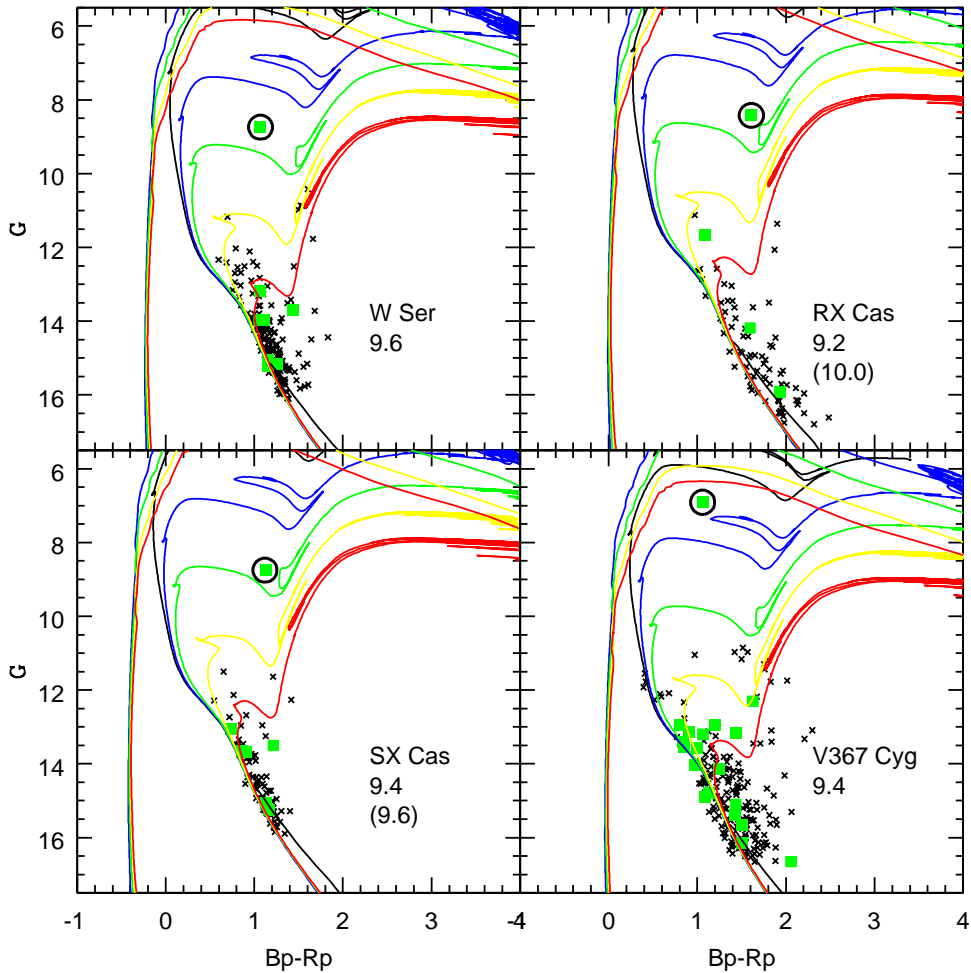}
\caption{Comparisons with non-rotating, solar-metallicity 
MIST isochrones. The sequences have ages log(t$_{yr}$) = 
7.5 (black), 8.0 (blue), 8.5 (green), 9.0 (yellow), and 
9.5 (red). Points are plotted using the same identification scheme as in 
Figure 6a. The log(t$_{yr}$) estimated for each system is shown under 
the system name. The entries in brackets for RX Cas and SX Cas are 
age estimates that are based on the second brightest main sequence star, as a 
SGB is not visible in the final sample CMDs of those systems. 
The estimated uncertainties in the ages are 0.1 -- 0.2 dex (see text).}
\end{figure}

\begin{figure}
\figurenum{7b}
\epsscale{0.9}
\plotone{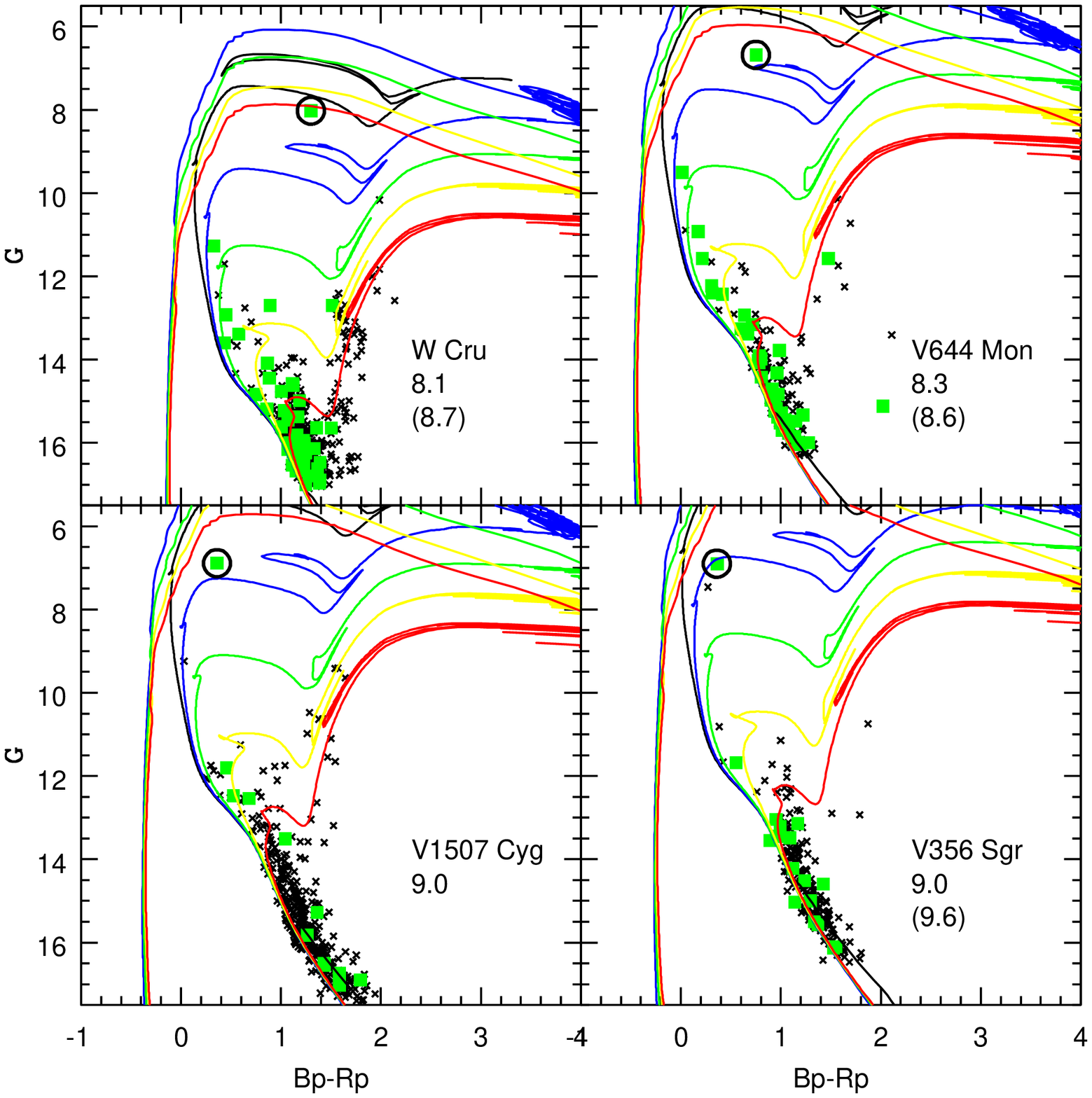}
\caption{Same as Figure 7a, but showing CMDs for W Cru, V644 Mon, 
V1507 Cyg, and V356 Sgr. Two age estimates are given for W Cru, V644 
Mon, and V356 Sgr, and the entry in brackets is that based on the 
second brightest main sequence star.} 
\end{figure}

	For four of the systems (W Ser, SX Cas, V367 Cyg, 
and V1507 Cyg) the age estimate is based on a match 
to both the MSTO and SGB. The final sample CMD of 
V367 Cyg has a well-defined MSTO region that includes 
a SGB, and the SGB is matched by the isochrone that is anchored on the 
brightest main sequence star. The age estimate for this system is thus 
considered to be the most secure. 

	There are no candidate SGB stars in the final samples of 
RX Cas, V644 Mon, and V356 Sgr to guide comparisons with 
the isochrones. W Cru is another system for which the location of the 
SGB may be problematic: while there are two stars that define an 
apparent SGB, there is also a blue star in the final sample that is 
$\sim 2$ magnitudes above the next brightest main sequence star in the CMD. 
For these four systems the age estimate immediately under the 
system name is based on the brightest main sequence star, 
while the age in brackets is based on the second 
brightest MS star. The second entry for these systems 
thus assumes that the brightest blue star is a blue 
straggler or an interloper from the field.
In the case of W Cru the isochrone that uses the second brightest 
main sequence star to define the MSTO is more-or-less consistent 
with the location of the two candidate SGB stars. 

	The W Ser systems tend to have absolute $G$ magnitudes that 
are at least 2 magnitudes brighter than 
the next brightest objects in the final 
samples. Of particular interest is that the absolute magnitudes and colors 
of W Ser, RX Cas, and SX Cas agree to within a few tenths of a magnitude, 
and so these systems may share similar intrinsic properties. Evidence to 
support this notion comes from the brightest main sequence stars in their 
final samples. The MSTO brightnesses of W Ser and 
SX Cas agree to within a few tenths of a magnitude, 
whereas the brightest blue star in the RX Cas final sample 
is $\sim 1$ magnitude brighter than in the final samples of W Ser and 
SX Cas. The absolute magnitude of the brightest main sequence star in 
RX Cas is not vastly different from that of W Ser and SX Cas, even though 
there are only three stars in the final sample of that system. 
To the extent that the three systems have similar intrinsic 
properties then a comparison of the age estimates for these systems 
based on the MSTO suggests an uncertainty in the log(t$_{yr}$) estimates 
of $\sim 0.1 - 0.2$ dex.

	Small number statistics contribute to uncertainties in 
the location of the MSTO for systems where the final sample 
is poorly populated. However, the comparison of the MSTO brightness 
estimates for W Ser, RX Cas, and SX Cas in the previous paragraph suggest that 
stochastic effects contribute only a 0.1 -- 0.2 dex error in log(t$_{yr}$). 
Contamination from even a single field star that has proper motions that are 
fortuitously similar to those of the W Ser systems might also bias the age 
estimates of those systems with only small numbers of stars 
in the final sample.

	A comparison between the isochrones and 
the baseline CMDs might provide guidance into 
field star contamination. In most cases the baseline sequences along 
the SGB are matched by log(t)=9.5 isochrones, and this is close to 
the age estimates for W Ser, RX Cas, SX Cas, V367 Cyg, and V356 Sgr. 
Even though the age estimate for V367 Cyg is consistent with that of the SGB 
population in the baseline sample, the CMD of the V367 Cyg final sample 
contains a well-defined SGB sequence that follows the isochrone that is 
consistent with the brightest candidate MSTO star. This suggests that 
the ages estimated from the isochrones for W Cru, V367 Cyg, V644 Mon, and 
V1507 Cyg are not skewed by field stars. While the evidence against possible 
contamination from one or two background stars is not as strong for 
W Ser and RX Cas, we note that both are near the outer fringes of the main 
data cloud on the proper motion diagram, reducing the chances of 
contamination from field stars.

\subsection{Estimating Initial Masses}

	The ages of the groupings and 
the masses of MSTO stars are listed in the second and third columns of 
Table 2. The uncertainties in the masses assume a $\pm 0.2$ dex error in 
log(t$_{yr}$) (see previous section). The uncertainties 
in the MSTO masses do not include systematic errors 
due to metallicity, reddening, etc. However, errors in these quantities are 
expected to introduce uncertainties in masses that are smaller than those 
that arise due to uncertainties in age and in stellar properties, such 
as the rate of rotation (see below).

	If the W Ser systems are experiencing Case B mass transfer then 
the donor star was likely on its first transit across the CMD when mass 
transfer was initiated. As the donor has then only 
recently evolved off of the main sequence then the {\it initial} mass of 
the donor star should not depart markedly from 
that of a star at the MSTO of the group or cluster 
in which it is a member (but see caveats below). Given the rapid pace of 
mass transfer during the initial stages of the interaction \citep[e.g.][]
{nelandegg2001}, then unless the system is viewed at a fortuitous 
time at the very beginning of mass transfer then the present-day mass of the 
donor star will be lower than its initial value.

\begin{deluxetable}{ccccc}
\tabletypesize{\scriptsize}
\tablecaption{Age and MSTO masses}
\tablehead{Name & log(Age)$_{0.0}$\tablenotemark{a} & MSTO Mass$_{0.0}$\tablenotemark{b} & Total System Mass$_{0.0}$\tablenotemark{c} & M2016 Mass\tablenotemark{d} \\
 & & (M$_{\odot}$) & (M$_{\odot}$) & (M$_{\odot}$) }
\startdata
W Ser & 9.6 & 1.1$^{+0.2}_{-0.1}$  & 2.2$^{+0.4}_{-0.2}$ & 2.5 \\
 & & & & \\
RX Cas & 9.2 & 1.5$^{+0.3}_{-0.2}$ & 3.0$^{+0.6}_{-0.4}$ & $7.6 \pm 0.6$ \\
 & (10.0) & (1.0$^{+0.1}_{-0.1}$) & (2.0$^{+0.2}_{-0.2}$) & \\
 & & & & \\
SX Cas & 9.4 & 1.3$^{+0.2}_{-0.2}$ & 2.6$^{+0.4}_{-0.4}$ & $6.6 \pm 0.6$ \\
 & (9.6) & (1.1$^{+0.2}_{-0.1}$) & (2.2$^{+0.4}_{-0.2}$) & \\
 & & & & \\
V367 Cyg & 9.4 & 1.3$^{+0.2}_{-0.2}$ & 2.6$^{+0.4}_{-0.4}$ & $7.3 \pm 1.0$ \\
 & & & & \\
W Cru & 8.1 & 4.0$^{+0.9}_{-0.7}$ & 8.0$^{+1.8}_{-1.4}$ & 9.0 \\
 & (8.7) & (2.3$^{+0.5}_{-0.3}$) & (4.6$^{+1.0}_{-0.6}$) & \\
 & & & & \\
V644 Mon & 8.3 & 3.3$^{+0.7}_{-0.5}$ & 6.6$^{+1.4}_{-1.0}$ & -- \\
 & (8.6) & (2.5$^{+0.6}_{-0.4}$) & (5.0$^{+1.2}_{-0.8}$) & \\
 & & & & \\
V1507 Cyg & 9.0 & 1.8$^{+0.3}_{-0.3}$ & 3.6$^{+0.6}_{-0.6}$ & -- \\
 & & & & \\
V356 Sgr & 9.0 & 1.8$^{+0.3}_{-0.3}$ & 3.6$^{+0.6}_{-0.6}$ & -- \\
 & (9.6) & (1.1$^{+0.2}_{-0.1}$) & (2.2$^{+0.4}_{-0.2}$) & \\
\enddata
\tablenotetext{a}{LLog of age in years as measured from non-rotating stellar 
models. The estimated uncertainties are 
0.1 -- 0.2 dex.}
\tablenotetext{b}{Mass of star at MSTO obtained from non-rotating stellar 
models. The quoted errors assume a $\pm 0.2$ dex uncertainty in 
log(t$_{yr}$) (see text).}
\tablenotetext{c}{Total system mass assuming an initial mass ratio of unity 
and conservative mass transfer.}
\tablenotetext{d}{Total system mass from Tables 4 and 5 of \citet{menetal2016}.}
\end{deluxetable}

	If the mass transfer is a conservative process 
then the mass lost from the donor star will remain in the system. Therefore, 
working under the assumption of conservative mass exchange then 
if an initial system mass ratio is assumed then the total system mass 
can be estimated knowing the initial mass of the donor star. 
Of course, the present-day mass ratio of these systems is likely very different 
from the initial value given that mass transfer 
is now underway. System mass estimates that assume an initial 
mass ratio of unity are listed in the fourth column of Table 2. 
These are upper limits to the system mass since (1) the donor star 
must have initially been the more massive star in the system 
(i.e. the mass ratio must have been less than unity), and (2) the mass 
transfer process is likely not conservative - at least some mass is 
almost certainly ejected from the system. 

	Some of these systems have mass estimates from other sources. 
\citet{menetal2016} compile mass estimates for the five 
systems identified by \citet{pk1978}, and these masses are 
listed in the last column of Table 2. These estimates are based on 
spectroscopic analyses of these systems, and the 
uncertainties are those listed in Table 5 of \citet{menetal2016}.  

	The mass estimates for W Ser systems that 
are based on photometric and spectroscopic 
measurements can be highly uncertain due to assumptions made when interpreting 
the origins of the signal, as light does not come solely 
from the component stars. One implication is that absorption and emission 
features may not track the orbital motions of the components. The range 
in mass estimates made for a system can be substantial. For example, 
\citet{dav2022a} compiles mass estimates for the components of V367 Cyg, and 
the various estimates for the total system mass fall between 7 M$_\odot$ 
and 33 M$_\odot$.

	Three of the systems do not have mass estimates in the final column 
of Table 2. We are not aware of modern mass estimates for V1507 Cyg in the 
published literature. As for V644 Mon and V356 Sgr, \citet{auf1994} finds 
component masses of 4 and 13 M$_{\odot}$ for the former, while \citet{fraetal1994}
find masses of 3 and 11 M$_{\odot}$ for the components of the latter. 
Details regarding how these masses were determined 
have not been published, and so we opt not to include them in Table 2. 
However, if these mass estimates are correct then V644 Mon and V356 Sgr 
are then the most massive systems in the sample examined here.

	With the exceptions of W Ser and W Cru, the initial system masses 
inferred from the MSTO and assuming a mass ratio of unity are much smaller 
than the masses listed by \citet{menetal2016}. 
The MSTO mass estimates in the third column of Table 2 are 
based on non-rotating stellar models. Rotation induced mixing delays 
evolution so that stars with a higher mass will maintain central hydrogen 
burning longer than non-rotating stars. 
In fact, a wide dispersion in rotational velocities has been detected 
in Galactic clusters with ages that are comparable to those of some of the 
groups found here \citep[e.g.][]{basetal2018} and also in intermediate age 
clusters in the Large Magellanic Cloud \citep[e.g.][]{macandbro2007}. 
If models that assume $\frac{v}{v_{Crit}} = 0.4$ are used to compute 
MSTO masses then the results in the fourth column of Table 2
increase by $\sim 0.2 - 0.3$M$_{\odot}$ ($\sim 0.4 - 0.6$M$_{\odot}$ 
for initial system masses), in the sense that higher offsets 
apply to the youngest systems. This is not sufficient to bridge the gap 
in masses listed in the fourth and fifth columns of Table 2. 
Therefore, while including moderate amounts of rotation 
will elevate mass estimates obtained from the 
MSTO of any cluster/groups, comparatively high rates of rotation 
among the MSTO stars are required to achieve better agreement with the 
\citet{menetal2016} mass estimates. 

	Of course, another possible (and perhaps more plausible)
explanation is that the donor stars in W Ser systems have had 
their evolution delayed by a mechanism other than rotation.
Indeed, as binary systems in which there has been mass 
transfer, the evolution of the component stars have likely 
departed from that expected for single stars that have the same initial 
masses \citep[e.g.][]{vananddeg2020, vanetal2011, lan2012, huretal2002}.
We also note that one system, V644 Mon, has been classified 
as a Be star \citep[][and references therein]{wel1979}. 
Many such objects are rapid rotators, possibly approaching 
the critical break-up velocity \citep[]{towetal2004}. 

\section{THE CIRCUMBINARY ENVIRONMENT}

	\citet{desetal2015} model the early stages of 
mass transfer in CBSs, including mass outflow 
that is powered by hot spots on the accretion disk that surrounds 
the secondary star. A circumsystem dust envelope 
forms as grains condense out of the ejected material, and the resulting 
envelope may extend out to many AU. \citet{desetal2015} 
simulate a system with intermediate mass components, and conclude  
that envelopes around systems at distances out to at least a few 
hundred parsecs should be observable in the MIR with a moderate-sized 
space-based telescope. In fact, envelopes like those simulated 
by \citet{desetal2015} are resolved around nearby W Ser stars. 
\citet{dav2022a} examined WISE W2 survey images in order to 
study the morphology of the envelope around the shell system V367 Cyg. 
In the present study, W2 images of all eight systems are examined. 

	The W Ser systems are the brightest sources in these images, and 
their light profiles swamp faint circumsystem 
emission. The signal from each system was removed by subtracting out 
a PSF that was constructed by median-combining 
normalized light from isolated stars. Bright and isolated PSF stars were not 
available in all of the image tiles, and so PSFs were 
constructed by including stars from many of the fields. 
This approach has the merit of averaging out structure in the 
PSF that is a result of the scanning technique used to record the raw images 
that are ultimately combined to form the final WISE images, 
coupled with variations in the PSF across the WISE image plane. \footnote[2]
{https://wise2.ipac.caltech.edu/docs/release/allsky/expsup/sec$4_4$c.html\#psf} 
The light near the center of the systems has contributions from both the 
binary system and a fainter envelope, and the PSF was scaled to minimize the 
residuals near the central regions of each W Ser system in PSF-subtracted 
images.

	PSF-subtracted W2 images are shown in Figure 8. 
The images in this figure are displayed with a common scaling factor to 
allow system-to-system variations in envelope brightness to be assessed. 
Envelopes are detected around W Ser, RX Cas, V367 Cyg, W Cru, and 
V644 Mon. The prominent central holes indicate that the PSF has been 
over-subtracted, despite scaling the PSF to minimize 
residuals. PSF-subtraction using a more refined estimate of 
the true brightness of the binary system should produce 
even more prominent envelopes than those seen in Figure 8.

\begin{figure}
\figurenum{8}
\plotone{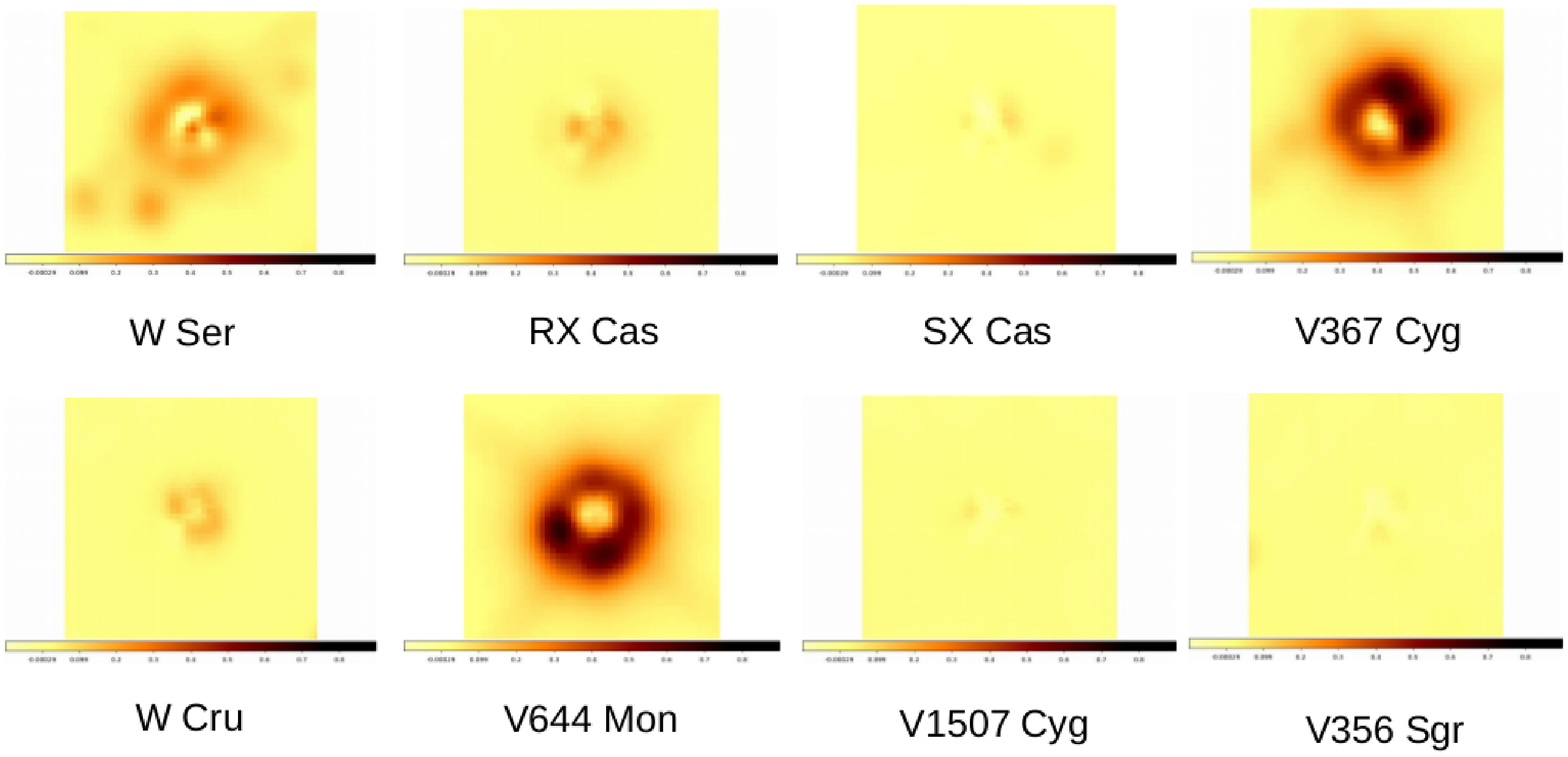}
\caption{Envelopes around the W Ser systems. Each panel shows 
an image in W2 that subtends $70 \times 70$ arcsec. A PSF 
that has been scaled to minimize the residuals near the center of each system 
has been subtracted from each image to highlight signal 
from envelopes and surrounding stars. The PSF-subtracted 
images are displayed with a common scaling factor to 
facilitate system-to-system comparisons of envelope brightness. 
Extended envelopes are clearly seen around W Ser, RX Cas, V367 Cyg, W Cru, 
and V644 Mon. While not evident in this figure, a very faint envelope 
is also present around V1507 Cyg. The prominent sub-structures 
within the envelopes are likely artifacts of the 
PSF-substraction process (see text).} 
\end{figure}

	W Cru is the most distant system and so might be expected to have 
the faintest and most compact envelope when compared with the other systems 
if it is in a similar stage of evolution. In fact, the envelope around W Cru is 
fainter and more compact than those around W Ser, V367 Cyg, and V644 Mon. 
Factors that are intrinsic to the W Cru system, such as the rate of mass 
transfer and the properties of the accretion 
disk hot spot, are also expected to play a role in defining the 
brightness and spatial extent of a circumsystem envelope. 

	While not apparent from Figure 8, a weak envelope is 
also seen around V1507 Cyg. This system is at a distance 
that is comparable to those that have more 
prominent envelopes, and so the faint emission in the W2 
images is not a distance effect. As for SX Cas and V356 Sgr, there is 
no convincing evidence for envelopes around these systems. It has been 
suggested that V356 Sgr is near the end of the rapid mass transfer phase of 
evolution \citep[]{dometal2005}, and so the dust envelope may be in the process 
of dissipating. We suspect that V1507 Cyg and SX Cas are also at an 
evolutionary stage in which a prominent envelope either has not yet formed, 
or is dispersing.

	The light distributions in Figure 8 
are not azimuthally symmetric, and obvious sub-structures 
are seen in the envelopes of all 5 of the systems in which a disk is 
evident. The most prominent such structures are in the envelopes 
around V367 Cyg and V644 Mon. We suspect that these are artifacts 
of variations in the WISE PSF that are inherent to the raw scans that were 
combined to make the final images. This can be checked with observations from 
another facility. However, as demonstrated in Figure 14 of \citet{dav2022a}, 
the FWHM of the PSF would have to be in error by almost a factor of two to 
account for the extended halos that are seen in Figure 8.

\section{COMMENTS ON INDIVIDUAL SYSTEMS}

\noindent{\bf W Ser:} The angular and radial distributions of the final 
and baseline samples of this system do not differ significantly, 
suggesting that W Ser is not in a recognizable asterism. 
Nevertheless, there are hints that it belongs to a stellar 
ensemble that is distinct from most of its neighbors. 
First, the kinematic properties of the final and baseline samples 
differ, in the sense that the location of W Ser on 
the proper motion diagram (Figure 1a) is displaced from 
the majority of its neighbors. The stars in the final sample 
are thus also displaced on the proper motion plane from the majority of 
stars in the baseline sample. Second, while the CMD of stars in the final 
sample is sparsely populated, the sequence defined by final sample stars, 
which includes a possible SGB star, passes below 
the log(t$_{yr}$) = 9.5 isochrone, which in turn is a few tenths of a dex older 
than would be infered for the baseline sample stars. 
It can be seen from Figure 8 that there are stars 
that are offset by only a few tens of arcsec ($\sim 0.1 - 0.2$ parsecs at 
the distance of W Ser) from the system. If subsequent 
observations find that these stars are at the same distance as W Ser then 
they may have affected the evolution of the system. While 
it is not clear if these are physically related to W Ser or are 
foreground/background field stars, stars with a similar instrinsic 
IR brightness are not seen near the other systems.
W Ser is one of only two systems studied here (the other is W Cru) in which the 
total system mass estimated from the MSTO agrees with 
a mass estimate based on other means. There is clear evidence 
for an extended envelope around W Ser in the mid-infrared, consistent 
with the notion that it is experiencing rapid mass transfer.

\vspace{0.3cm}
\noindent{\bf RX Cas:} GAIA parallaxes suggest that RX Cas is the closest 
of the eight systems studied here, and so the final and baseline samples are 
extracted from the smallest spatial volume. It is then perhaps not surprising 
that the baseline sample for this system has the smallest number of objects. 
Like W Ser, the location of RX Cas on the proper motion diagram is offset 
from the center of the data cloud, indicating that it and the 
final sample stars have kinematic properties that differ from those of 
many of its neighbors. Unfortunately, the final sample has only 3 stars aside 
from RX Cas itself, and this frustrates efforts to examine the structural 
properties of the final sample stars and estimate the initial mass of the 
donor. Still, the angular distribution of final sample stars in Figure 3a is 
lop-sided when compared with that of the baseline sample, and this result 
is significant at the 85\% confidence level. While not conclusive, 
it suggests that if RX Cas is accompanied by stars that formed 
with it then they are now appear to be part of a highly diffuse 
stream-like moving group. A search for other members of a 
moving group associated with RX Cas might then prove to be rewarding.
There is evidence for a circumsystem envelope in the W2 images.

\vspace{0.3cm}
\noindent{\bf SX Cas:} SX Cas is unique among the systems studied here 
in that the radial distribution of stars in the final sample is 
more compact than that of stars in the baseline sample, 
and this is significant at the 95\% confidence level. This is consistent 
with SX Cas being at or near the center of a cluster, lending confidence 
to the notion that the final sample stars are physically related to SX Cas. 
The brightest candidate MS star in the final sample is 1.5 mag fainter than the 
brightest MS star in the baseline sample. There is also a possible SGB star 
in the final sample, which has the potential to provide leverage for 
comparisons with isochrones. The CMD of the final sample stars is consistent 
with log(t$_{yr}$) = 9.4 or 9.6, depending on the weight 
assigned to the SGB star when making comparisons with the isochrones. 
There is no evidence for an extended envelope in W2, suggesting 
that SX Cas is in an evolutionary stage that differs from that of most 
of the other systems. Still, the integrated 
photometric properties of SX Cas based on the GAIA photometry are 
very similar to those of W Ser and RX Cas, both of which have extended 
envelopes.

\vspace{0.3cm}
\noindent{\bf V367 Cyg:} This system is located in a part of the sky where 
there is active and recent star formation. However, any association with 
active star-forming regions is likely a projection effect. 
While V367 Cyg is close to the young cluster M29 on the sky, the parallaxes 
of the brightest objects in M29 are roughly one half that of V367 Cyg. 
However, V367 Cyg may still be behind some of the dust that obscures 
this part of the sky. The angular and radial distributions of stars in the 
baseline and final samples of V367 Cyg are not significantly different, 
and the angular distribution of both samples are indicative of 
a lop-sided distribution. We suspect that this is due to foreground extinction. 
The CMD of final sample stars shows a well-defined MSTO complete 
with a SGB, and the match between the final sample CMD and isochrones 
is likely the most secure of the systems considered here. 
The CMD of the final sample stars are matched with a 
log(t$_{yr}$) = 9.4 solar metallicity isochrone, and the 
corresponding MSTO mass is 1.3M$_{\odot}$. \citet{dav2022a} 
reviewed mass estimates for V367 Cyg. Those that include a disk 
when modelling light curves, and hence presumably have realistic input 
physics, yield masses of 3.3 and 4.0 M$_{\odot}$ 
\citep[]{zolandogl2001}, and these were adopted by \citet{menetal2016}. 
The reliable comparison with isochrones notwithstanding, 
the system mass estimated from the MSTO and assuming a mass ratio of 
unity is much lower than previously published values.
As noted by \citet{dav2022a}, there is a prominent circumsystem envelope 
visible in the MIR. This may be related to the shell of material that blankets 
the system spectrum \citep[e.g.][and references therein]{hacetal1984}.

\vspace{0.3cm}
\noindent{\bf W Cru:} W Cru is the most distant system considered in this 
paper, and so the final and baseline samples are extracted from a larger 
volume of space than is the case for the other systems. It is then not 
surprising that W Cru has the largest baseline sample, although this also 
means that the final sample is then subject to 
the highest fractional contamination from field stars. 
That physically unrelated stellar groupings along the line of sight that have 
fortuitously similar proper motions might be sampled is of particular concern 
in richly populated low Galactic latitude fields such as those in Crux. 
In fact, the location of W Cru on the proper motion diagram (Figure 1b) 
is near the main concentration of objects in the baseline sample. Still, 
there is some evidence (albeit marginal) for structural differences between 
the final and baseline samples, in the sense that the radial distributions 
of the two samples differ at the 70\% confidence level. 
More significantly, the CMDs of the final and baseline samples clearly 
differ. The CMD of the baseline sample has a morphology 
that is reminiscent of the CMDs of field stars \citep[e.g.][]{babetal2018}, 
and comparisons with solar metallicity isochrones yield log(t$_{yr}$) = 9.5 
for stars on the SGB. In contrast, comparisons with isochrones suggest 
that the stars in the final sample have an age log(t$_{yr}$) = 8.1 or 8.7, 
depending on the weights assigned to the brightest candidate main sequence 
star and the possible SGB star that has Gp $\sim 12.5$ and Bp--Rp $\sim 1.4$. 
If the younger age estimate is adopted then the total system mass based on 
the MSTO is close to that found in other studies; in this case 
the SGB star is a field interloper. The mass estimates would be in even better 
agreement if the stars near the MSTO in the final sample CMD evolved with 
moderate amounts of rotation. The donor star in W Cru has 
the highest initial mass among the systems considered in 
this paper, and this is perhaps not surprising given the long orbital period, 
the spectral type of the donor star (late F to early G supergiant), and the 
absolute $G$ brightness of the system. The 
W2 image of W Cru indicates that an envelope is present, so the 
system is ejecting mass into the surrounding ISM. 

\vspace{0.3cm}
\noindent{\bf V644 Mon:} While the projected location of V644 Mon 
on the sky is close to CMa OB1, the GAIA parallax places it behind 
this association and related sub-structures \citep[]{sanetal2021}. 
Both the baseline and final samples have a lop-sided projected 
on-sky distribution, and we suggest that this is likely due to 
non-uniform foreground extinction along the line of sight, 
much of which may be related to CMa OB1. While the V644 Mon final sample is 
well-populated, there is no evidence for structural differences in the 
angular and radial distributions of stars in the final and baseline samples. 
Thus, any stellar grouping associated with V644 Mon that contains stars in the 
current dataset does not form an asterism that is distinct from its immediate 
neighbors. As for the age of V644 Mon, there is a 
possible SGB star in the final sample, and two ages have been estimated 
depending on whether or not the red object is matched by the isochrones. 
Age estimates that differ by 0.3 dex are found, and these produce 
MSTO mass estimates that differ by 0.4M$_{\odot}$. 
Both estimates predict super-solar initial masses for the 
donor star, as might be expected given the Be classification \citep{wel1979}. 
The donor star in V644 Mon is the second most massive such object in our 
sample, and the system has the second highest absolute $G$ magnitude based 
on the entries in Table 1. The two systems with the most massive donors 
(W Cru and V644 Mon) thus are also the most luminous at visible wavelengths.
This is noteworthy as the method used to estimate ages and masses for these 
systems does not rely on their intrinsic photometric properties, but instead 
on those of companion stars. There is evidence for strong, resolved 
circumsystem emission in the MIR, as might be expected given the 
Be classification.

\vspace{0.3cm}
\noindent{\bf V1507 Cyg:} The location of V1507 Cyg on the sky places it 
close to previously identified clusters. It is roughly 50 arcmin from the 
moderately young and compact open cluster NGC 6834, although that cluster 
has a parallax $\pi \sim 0.3$ mas, and so is much more distant than V1507 Cyg. 
Turner 9 is another seemingly nearby cluster, but again it is more distant, 
with $\pi \sim 0.6$ \citep[]{canandand2020}. Still, there is some 
evidence for clustering as the proper motion diagram of the baseline sample 
shows a concentration of objects with $\mu_\delta \sim -4$ mas/year 
and $\mu_\alpha \sim -1$ mas/year that is offset from 
V1507 Cyg in this diagram. The majority of the stars in this concentration 
are not included in the final sample given the extraction 
criterion described in Section 3. Nonetheless, there is 
an obvious lop-sided distribution of final sample stars in Figure 3b, 
in the sense that the angular distribution of the final sample 
differs from that of the baseline sample at the 70\% confidence level, 
while the radial distributions of stars in the two samples differ at the 
75\% confidence level. The former result is consistent with V1507 Cyg 
being on the periphery of a cluster or moving group. 
However, the CMD of the baseline sample has a prominent SGB, as 
expected if there is a large field star population, while there are 
no candidate SGB stars in the final sample. The MSTO of the final sample, 
which is close to the brightest main sequence star in the baseline 
sample, is consistent with a log(t$_{yr}$) = 9.0 solar metallicity 
isochrone, which is $\sim 0.5$ dex younger than the stars on the SGB 
in the baseline sample. Circumstellar emission is weak 
in W2 when compared with the other W Ser systems, but is still present.

\vspace{0.3cm}
\noindent{\bf V356 Sgr:} V356 Sgr is in a crowded field, and neither 
the angular nor radial distributions of objects on the sky in the 
final sample differ significantly from that of 
the baseline sample. There is a $\sim 1.5$ 
magnitude gap between the two brightest MS stars in the 
final sample, leading to considerable ambiguity in the age estimate. 
Indeed, depending on the adopted brightness of the MSTO then the CMD 
is consistent with non-rotating solar metallicity isochrones having 
ages log(t$_{yr}$) = 9 or 9.6. Despite being one of the closest 
systems, there is no evidence for extended circumsystem emission. 
\citet{dometal2005} suggest that V356 Sgr may be 
near the end of the rapid mass transfer phase. 
If this is the case then if a circumsystem shell that was comparable 
to those in the other systems formed earlier then it 
appears to have dissipated.
 
\section{DISCUSSION \& SUMMARY}

	The W Ser systems are interacting binaries that are likely 
experiencing the rapid phase of mass transfer after the more massive star 
has evolved to fill its Roche lobe. The masses and ages of the stars in these 
systems are highly uncertain given the evolved state of one component, which 
makes it much brighter than its companion, coupled with 
other sources of light that are not directly related to the 
components and the presence of an obscuring disk around the gainer. If 
stars that formed at the same time as these systems can be 
identified then it may be possible to estimate their ages, and then set 
limits on the initial mass of the donor star that has (presumably) recently 
evolved off of the main sequence. This information is essential to forecast 
the future evolution of the W Ser systems and place them in the context of CBSs 
that are in more advanced stages of evolution. 

	In this study, astrometric and photometric measurements extracted 
from the GAIA DR3 have been used to identify stars that are associated with 
five classical W Ser systems and three related objects. 
A volume with a 10 parsec radius on the sky and a depth along the line 
of sight that depends on the system distance has been 
searched around each system to identify possible companions. The 
10 pc radius that was adopted for the projected 
coverage on the sky was selected to sample a 
moderately diffuse, intermediate age cluster like the Hyades, and so is a 
conservative starting point for a search for companions. \citet{meietal2021} 
find that the coronae of clusters in the age range 30 -- 300 Myr 
may extend over distances of 100 pc. However, such coronae are very 
diffuse, and it is not clear if these would survive over timescales 
in excess of a Gyr, which we find to be the age for many of the 
stellar ensembles associated with the W Ser systems. Thus, we prefer to 
use the more conservative 10 pc search radius.

	Two samples of stars have been identified for each W Ser system: 
one that includes all of the stars within the search volume after 
filtering based on the uncertainties in the 
parallax measurements (the baseline sample), and a subset of 
the baseline sample that has been subject to additional filtering based 
on proper motions (the final sample). The stars that have 
distances and kinematic properties that are similar to 
those of the W Ser systems within this volume are likely fossil 
remnants of what was once a much larger star cluster. 
The CMDs of these groupings have been used to estimate ages 
and initial donor star masses. In some cases, stars in the final sample 
define an asterism that stands out from the projected distribution of 
other neighboring stars. 

	The strengths and morphologies of circumsystem 
envelopes in the mid-infrared have also been examined using 
images from the WISE survey. Evidence for extended envelopes is found 
around most of the systems. System-to-system variation in 
the strength of this emission is seen, and this is interpreted in the context 
of differing stages in the evolution of the systems. 

\subsection{Parallaxes, Proper Motions, and Location on the Sky}

	Parallaxes are of basic importance for selecting companions of 
the W Ser systems, and these measurements are likely subject to systematic 
errors \citep[e.g.][]{linetal2021}. While \citet{staandtor2021} 
and \citet{renetal2021} describe how these systematics have been reduced 
in progressive GAIA data releases using eclipsing binaries 
as calibrators, systematics still remain. 
In this study the range of parallaxes that have been searched 
for companions around each system has been defined to take into account 
possible systematics that arise due to differences in source brightness. 
In particular, the range of parallaxes considered for each system is based on 
the uncertainties in the parallax measurements of that system. This was done 
because the uncertainties in the parallaxes of 
the W Ser systems tend to be comparable to the possible systematic errors due 
to source brightness. Restricting the search volume according to the 
uncertainty in the parallax thus allows at least some fraction of 
actual cluster members to be selected even if systematic errors due to 
source brightness are present. 

	An unfortunate consequence of the use of the 
uncertainty in parallax rather than a fixed search depth 
is that the search volume increases when systems at progressively larger 
distance (and hence larger uncertainties in distance) are considered.
However, if a fixed physical depth (say 10 pc, corresponding to 
the projected search radius on the sky) had been employed instead 
then the range in parallaxes that are searched would have been smaller than 
that resulting from the uncertainties in the system parallax, with the results 
that actual cluster members might be missed if there are systematic 
errors due to source brightness or if the random uncertainties in parallax 
are large.

	Further mitigation of systematic errors in 
parallax was provided with other selection criteria. First, 
the projected area sampled on the sky around each W Ser 
system is less than a degree in radius, thereby reducing potential 
systematic errors due to location on the sky. Second, upper limits have 
been placed on random uncertainties in the parallaxes. This is an 
obvious criterion to limit the number of sources that physically fall 
outside the volume of interest but may creep into the sample because of errors 
in their parallax. It also sets a faint limit in target brightness, 
thereby reducing systematic uncertainties in parallaxes that might arise 
due to source brightness. 

	Proper motions are also critical for the identification of 
possible companions. Obvious groupings of objects in the proper motion 
plane are seen around V367 Cyg, W Cru, and V644 Mon. However, similar 
concentrations are less obvious for the other systems when they are considered 
individually. An extraction criterion for proper motions was thus found by 
combining the proper motion diagrams of four of the closest 
W Ser systems after adjusting for system-to-system differences in 
proper motion. Clear evidence of clustering in the proper motion plane was 
found, and the objects in these proper motion groupings are distributed over 
projected spatial scales on the sky of many parsecs. This indicates a tendency 
for the W Ser systems to be accompanied by a diffusely distributed group 
of companions, albeit in small numbers in some cases.  

	The kinematic properties of systems provide clues into 
their past evolution, as well as their relationship to other 
neighboring stars. The baseline sample is expected to 
contain a diverse mix of stars that spans a range of ages and 
kinematic properties, and so provides a reference sample for identifying 
structures that may be present in the final sample. 
W Ser and RX Cas are offset on the proper motion diagram, 
in the sense that they are located near the edges of the 
data distributions defined by the baseline samples. This suggests that 
the kinematic histories of these systems (as well as any accompanying stars) 
are distinct from those of the majority of stars in the baseline samples. 
In contrast, SX Cas, W Cru, and V644 Mon are more-or-less centered on 
the distribution defined by baseline stars in the proper motion plane.

	The projected distribution of stars on the sky in the final and 
baseline samples provides insights into the environment around the systems. 
The radial distribution of stars in the final sample of SX Cas differs 
from that of the baseline sample at the 95\% 
confidence level, and this is consistent with that system being near the 
center of an organized (but diffuse) asterism. 
Preliminary (which we interpret as a confidence level between 70 and 95\%) 
evidence is also found for components that have a distribution that is 
distinct from the baseline samples around RX Cas, W Cru, and V1507 Cyg. 
In the cases of RX Cas and V1507 Cyg the distribution of final sample stars 
on the sky hints that the W Ser system may be near the edge of a cluster or 
moving group. While there might then be merit to expanding the 
search area on the sky, we are hesitant to do so with the 
existing data as this increases the risk of contaminants 
from field stars that have fortuitously similar proper motions 
to those of the W Ser systems. 

	Despite efforts to suppress the presence of contaminants 
using proper motion as a filter, the final 
samples almost certainly contain sources that are not related to the W Ser 
systems. The final sample for W Cru is probably the most susceptible to 
such contamination, as the baseline sample is extracted 
over a distance of a few hundred pc along the 
line of sight. However, even in this extreme case 
there are hints that contamination from field stars is modest. 
For example, the distribution of objects on the sky in the baseline and 
final samples of W Cru differ, albeit at the 70\% confidence level. 
Moreover, the CMD of the W Cru final sample stars 
differs from that of the baseline population. 
The reduction of the uncertainties in the parallaxes combined with radial 
velocity measurements will provide information for confirming the 
identity of potential cluster members. 

\subsection{Ages and Masses}

	The locations of the MSTO in the CMDs of the final samples 
suggest that only V644 Mon and W Cru have ages that are less than 1 
Gyr, although those systems still have ages in excess of 100 Myr. 
The W Ser systems studied here thus appear to be 
intermediate age objects - they are likely not massive, young systems.
With the caveat that rotation and tidal effects may impact 
the pace of stellar evolution, the MSTO mass found from the CMDs of 
the final samples should be a close approximation to the {\it initial} 
mass of the donor star, as that star has presumably 
recently evolved off of the main sequence. If mass transfer has 
been more-or-less conservative up to the current epoch then a 
total system mass estimate can be made using the masses of MSTO stars 
in the final samples (column 4 of Table 2) after assuming an initial mass ratio.
The total initial system mass calculated in this way is an upper limit if 
an initial mass ratio of unity is adopted.

	The total initial mass estimates made from the CMDs of the final sample 
stars are lower than those found from other means 
(column 5 of Table 2) in 3 of 5 cases. What is the cause 
of the discrepancy between the total mass estimates 
for RX Cas, SX Cas, and V367 Cyg in Table 2? To begin,
mass estimates made from the combined analysis of spectroscopic and 
photometric information are not free of uncertainties. For example, V367 Cyg, 
has the best defined final sample CMD, and so has the 
greatest potential for achieving a reliable match with isochrones, and hence 
a reliable MSTO mass. Mass estimates for that system that are 
based on photometric and spectroscopic observations were reviewed by 
\cite{dav2022a}, who found a large range of values, none of which match those 
in the fourth column of Table 2.

	The brightness of the MSTO may have been 
underestimated in the final samples of RX Cas, SX Cas, and V367 Cyg. 
This might be expected given that only a modest number of final sample stars 
are present, and stochastic effects could then make finding a star 
at or near the 'true'  MSTO problematic. With only three stars in the 
final sample, the total mass estimate for RX Cas is the most vulnerable to 
stochastic effects. However, the actual MSTO on the CMD would have to be 
$\sim 2$ magnitudes brighter than found here to relieve the system mass 
discrepancy that is evident in Table 2. Therefore, even if the brightest 
blue star in the RX Cas baseline sample were used to 
estimate the system mass of RX Cas then there would still be a 
substantial mass discrepancy. As for V367 Cyg, the final sample CMD of 
that system is well-defined near the MSTO and SGB, and the 
discrepancy between the donor star mass estimates in Table 2 for this system 
is typical of those for the other two systems. While we are unable to rule 
out stochastic effects as a cause of the mass discrepancies for RX Cas and 
SX Cas, it is unlikely that these can account for all of the 
differences between the mass estimates of these systems in the fourth and 
fifth columns of Table 2.

	Internal mixing will prolong the lifetime of a star, 
and thereby cause a departure from mass estimates that are made from 
star models with no internal mixing, such as those applied here to obtain the 
MSTO masses that in turn serve as proxies for the initial masses of the 
donor stars. If the donor stars and the MSTO stars have had vastly different 
internal mixing histories then this could cause a discrepancy in the 
total mass estimates. That being said, examining models with moderate 
amounts of rotation suggest that the change in initial mass with 
respect to the non-rotating case in the mass range considered 
here (i.e. just above solar to a few times solar) is at most a few tenths 
M$_{\odot}$. To produce a large mass discrepancy then 
the donor stars in the CBSs must have experienced a much 
greater amount of mixing throughout their lives prior to the onset 
of interactions with their companions than 
the (presumably) isolated single stars near the MSTO. 
It would be of interest to obtain spectra of stars near the MSTO 
in the final samples to determine if these stars are fast rotators at 
the present day.

\subsection{Circumsystem Emission}

	If these systems are experiencing rapid mass transfer 
then models suggest that there will be a significant 
outward mass flow, producing circumsystem 
envelopes. There is well-established spectroscopic evidence for 
a circumsystem shell around V367 Cyg \citep[][and references therein]{dav2022a} 
and other classical W Ser systems \citep[e.g.][]{broetal2021}. Thermal 
emission will presumably originate in the outermost regions of the 
shells where grain formation is possible. In Section 7 evidence is presented 
for resolved envelopes around six of the systems in the MIR. It appears 
that the strength of the emission in the W2 filter is not correlated with 
the initial system mass estimates made in Section 5.2. While the most 
prominent detections are those around V367 Cyg and V644 Mon, 
these systems have very different initial mass estimates. 
It is more likely that the MIR emission is related 
to the rate of mass transfer, as would be consistent with 
the \citet{desetal2015} outflow model. We have spectra of V644 Mon in hand 
that cover much of its orbital cycle, and in a future paper we will use 
these to examine the properties of the circumsystem shell and rate of 
mass transfer.

\acknowledgements{
It is a pleasure to thank the anonymous reviewer for suggestions that greatly 
improved the manuscript. This research has made use of the NASA/IPAC Infrared 
Science Archive (https://doi.org/10.26131/irsa1), which is funded by the 
National Aeronautics and Space Administration 
and operated by the California Institute of Technology.
This work has also made use of data from the European Space Agency (ESA) mission
{\it Gaia} (\url{https://www.cosmos.esa.int/gaia}), processed by the {\it Gaia}
Data Processing and Analysis Consortium (DPAC,
\url{https://www.cosmos.esa.int/web/gaia/dpac/consortium}). Funding for the DPAC
has been provided by national institutions, in particular the institutions
participating in the {\it Gaia} Multilateral Agreement.
}


\parindent=0.0cm

\begin{thebibliography}{}

\bibitem[Augdenberg (1994)]{auf1994}
Aufdenberg, J. P. 1994, \aas, 185, 2110

\bibitem[Babusiaux et al. (2018)]{babetal2018}
Babusiaux, C., van Leeuwen, F., Barstow, M. A., et al. 2018, \aap, 616, A10

\bibitem[Bastian et al. (2018)]{basetal2018}
Bastian, N., Kamann, S., Cabrera-Ziri, I., Georgy, C., Ekstrom, S., Charbonnel, C., de Juan Ovelar, M., \& Usher, C. 2018, \mnras, 480, 3739

\bibitem[Bastian (2019)]{bas2019}Bastian, U. 2019, \aap, 630, L8

\bibitem[Broz et al. (2021)]{broetal2021}Broz, M., Mourard, D., Budaj, J. et al.
2021, \aap, 645, A51

\bibitem[Cantat-Gaudin \& Anders (2020)]{canandand2020}
Cantat-Gaudin, T., \& Anders, F. 2020, \aap, 633, 99

\bibitem[Choi et al. (2016)]{choetal2016}
Choi, J., Dotter, A., Conroy, C. et al. 2016, \apj, 823, 102

\bibitem[Davidge (2022a)]{dav2022a}
Davidge, T. J. 2022a, \aj, 164, 149

\bibitem[Davidge (2022b)]{dav2022b}
Davidge, T. J. 2022b, RNAAS, 6, 175

\bibitem[Davidge\& Milone (1984)]{dm1984}
Davidge, T. J., \& Milone, E. F. 1984, \apjs, 55, 571

\bibitem[Deschamps et al. (2015)]{desetal2015}
Deschamps, R., Braun, K., Jurissen, A., Siess, L., Baes, M., \& Camps, P. 2015, \aap, 577, A55

\bibitem[Dominis et al. (2005)]{dometal2005}
Dominis, D., Mimica, P., Pavlovski, K., \& Tamajo, E. 2005, \apss, 296, 189

\bibitem[Dotter (2016)]{dotetal2016}
Dotter, A. 2016, \apjs, 222, 8

\bibitem[Fraser et al. (1994)]{fraetal1994}
Fraser, K. L., Roby, S. W., \& Polidan, R. S. 1994, \baas, 29, 1287

\bibitem[Gaia Collaboration (2016)]{gai2016}
GAIA Collaboration: Prusti, T., de Bruijne, J. H. J., Brown, A. et al. 2016, \aap, 595, A1

\bibitem[GAIA Collaboration (2022)]{gai2022}
GAIA Collaboration, Vallenari, A., Brown, A. G. A., Prusti, T., et al. 2022, astro-ph, 2208.00211

\bibitem[Hack et al. (1984)]{hacetal1984}
Hack, M., Engin, S., \& Yilmaz, N. 1984, \aap, 131, 147

\bibitem[Halbedel (1989)]{hal1989}
Halbedel, E. M. 1989, \pasp, 101, 995

\bibitem[Hurley et al. (2002)]{huretal2002}
Hurley, J. R., Tout, C. A., Pols, O. A. 2002, \mnras, 329, 897

\bibitem[Jordi et al. (2010)]{joretal2010}
Jordi, C., Gebran, M., Carrasco, J. M., et al. 2010, \aap, 523, A48

\bibitem[Langer (2012)]{lan2012}
Langer, N. 2012, \araa, 50, 107

\bibitem[Lindgren et al. (2021)]{linetal2021}
Lindegren, L., Bastian, U., Biermann, M., et al. 2021, \aap, 649, A4

\bibitem[Mackey \& Brodie Neilson (2007)]{macandbro2007}
Mackey, A. D., \& Broby Neilson, P. 2007, \mnras, 379, 151

\bibitem[Maeder \& Meynet (2000)]{maeandmey2000}
Maeder, A., \& Meynet, G. 2000, \araa, 38, 143

\bibitem[Meingast et al. (2021)]{meietal2021}
Meingast, S., Alves, J., Rottensteiner, A. 2021, \aap, 645, A84

\bibitem[Mennickent et al. (2016)]{menetal2016}
Mennickent, R. E., Otero, S., \& Kolaczkowski, Z. 2016, \mnras, 455, 1728
 
\bibitem[Muzic et al. (2019)]{muzetal2019}
Muzic, K., Scholz, S., Pena Ramirez, K., et al. 2019, \apj, 881, 79

\bibitem[Nelson \& Eggleton (2001)]{nelandegg2001}
Nelson, C. A., \& Eggleton, P. P. 2001, \apj, 552, 664

\bibitem[Paxton et al. (2011)]{paxetal2011}
Paxton, B., Bildsten, L., Dotter, A. et al. 2011, \apjs, 192, 3

\bibitem[Paxton et al. (2013)]{paxetal2013}
Paxton, B., Cantiello, M., Arras, P. et al. 2013, \apjs, 208, 4

\bibitem[Paxton et al. (2015)]{paxetal2015}
Paxton, B., Marchant, P., Schwab, J. et al. 2015, \apjs, 220, 15

\bibitem[Plavec \& Koch (1978)]{pk1978}
Plavec, M., \& Koch, R. H. 1978, IBVS, No. 1482

\bibitem[Ren et al. (2021)]{renetal2021}
Ren, F., Chen, X., Zhang, H., de Grijs, R., Deng, L., \& Huang, Y. 2021, \apjl, 911, L20

\bibitem[Santos-Silva et al. (2021)]{sanetal2021}
Santos-Silva, T., Perottoni, H. D., Ameida-Fernandes, F., et al. 2021, \mnras, 508, 1033

\bibitem[Song et al. (2013)]{sonetal2013}
Song, H. F., Maeder, A., Meynet, G., Huang, R. Q., Ekstrom, S., \& Granada, A. 2013, \aap, 556, A100

\bibitem[Stassun \& Torres (2021)]{staandtor2021}
Stassun, K. G., \& Torres, G. 2021, \apj, 907, L33

\bibitem[Van Rensbergen \& De Greve (2020)]{vananddeg2020}
Van Rensbergen, W., \& De Greve, J. P. 2020, \aap, 642, A183

\bibitem[Van Rensbergen et al. (2011)]{vanetal2011}
Van Rensbergen, W., De Greve, J. P., Mennekens, N., Jansen, K., \& De Loore, C. 2011, \aap, 528, A16

\bibitem[Townsend et al. (2004)]{towetal2004}
Townsend, R. H. D., Owocki, S. P., \& Howarth, I. D. 2004, \mnras, 350, 189

\bibitem[Welin (1979)]{wel1979}
Welin, G. 1979, \aap, 79, 334

\bibitem[Wilson et al. (1984)]{wiletal1984}
Wilson, R. E., Rafert, J., \& Markworth, N. L. 1984, IAPPP Comm., 16

\bibitem[Wright et al. (2010)]{wrietal2010}
Wright, E. L., Eisenhart, P. R. M., Mainzer, A. K., et al. 2010, \aj, 140, 1868

@misc{https://doi.org/10.26131/irsa1,
  doi = {10.26131/IRSA1},
  url = {https://catcopy.ipac.caltech.edu/dois/doi.php?id=10.26131/IRSA1},
  author = {{Wright,  Edward L.; Eisenhardt,  Peter R. M.; Mainzer,  Amy K.; Ressler,  Michael E.; Cutri,  Roc M.; Jarrett,  Thomas; Kirkpatrick,  J. Davy; Padgett,  Deborah; McMillan,  Robert S.; Skrutskie,  Michael; Stanford,  S. A.; Cohen,  Martin; Walker,  Russell G.; Mather,  John C.; Leisawitz,  David; Gautier,  Thomas N.,  III; McLean,  Ian; Benford,  Dominic; Lonsdale,  Carol J.; Blain,  Andrew; Mendez,  Bryan; Irace,  William R.; Duval,  Valerie; Liu,  Fengchuan; Royer,  Don; Heinrichsen,  Ingolf; Howard,  Joan; Shannon,  Mark; Kendall,  Martha; Walsh,  Amy L.; Larsen,  Mark; Cardon,  Joel G.; Schick,  Scott; Schwalm,  Mark; Abid,  Mohamed; Fabinsky,  Beth; Naes,  Larry; Tsai,  ChaoWei}},
  title = {AllWISE Source Catalog},
  publisher = {IPAC},
  year = {2019}
}

\bibitem[Zola \& Ogloza (2001)]{zolandogl2001}
Zola, S., \& Ogloza, W. 2001, \aap, 368, 932

\end{thebibliography}
\end{document}